\documentclass[fleqn,usenatbib,useAMS]{mnras}

\usepackage{newtxtext,newtxmath}

\usepackage[T1]{fontenc}

\DeclareRobustCommand{\VAN}[3]{#2}
\let\VANthebibliography\thebibliography
\def\thebibliography{\DeclareRobustCommand{\VAN}[3]{##3}\VANthebibliography}

\usepackage{graphicx}	
\usepackage{amsmath}	
\usepackage{mathtools}
\usepackage{physics}
\usepackage{xcolor}

\def\setsymbol#1#2{\expandafter\def\csname #1\endcsname{#2}}
\def\getsymbol#1{\csname #1\endcsname}

\def\Planck{\textit{Planck}}

\newbox\tablebox    \newdimen\tablewidth
\def\leaderfil{\leaders\hbox to 5pt{\hss.\hss}\hfil}
\def\tablenote#1 #2\par{\begingroup \parindent=0.8em
    \abovedisplayshortskip=0pt\belowdisplayshortskip=0pt
    \noindent
    $$\hss\vbox{\hsize\tablewidth \hangindent=\parindent \hangafter=1 \noindent
    \hbox to \parindent{$^#1$\hss}\strut#2\strut\par}\hss$$
    \endgroup}

\def\L2{\ifmmode L_2\else $L_2$\fi}

\def\DeltaT{\ifmmode \Delta T\else $\Delta T$\fi}
\def\deltat{\ifmmode \Delta t\else $\Delta t$\fi}
\def\fknee{\ifmmode f_{\rm knee}\else $f_{\rm knee}$\fi}
\def\Fmax{\ifmmode F_{\rm max}\else $F_{\rm max}$\fi}
\def\solar{\ifmmode{\rm M}_{\mathord\odot}\else${\rm M}_{\mathord\odot}$\fi}
\def\Msolar{\ifmmode{\rm M}_{\mathord\odot}\else${\rm M}_{\mathord\odot}$\fi}
\def\Lsolar{\ifmmode{\rm L}_{\mathord\odot}\else${\rm L}_{\mathord\odot}$\fi}
\def\inv{\ifmmode^{-1}\else$^{-1}$\fi}
\def\mo{\ifmmode^{-1}\else$^{-1}$\fi}
\def\sup#1{\ifmmode ^{\rm #1}\else $^{\rm #1}$\fi}
\def\expo#1{\ifmmode \times 10^{#1}\else $\times 10^{#1}$\fi}
\def\,{\thinspace}
\def\lsim{\mathrel{\raise .4ex\hbox{\rlap{$<$}\lower 1.2ex\hbox{$\sim$}}}}
\def\gsim{\mathrel{\raise .4ex\hbox{\rlap{$>$}\lower 1.2ex\hbox{$\sim$}}}}

\def\simprop{\mathrel{\raise .4ex\hbox{\rlap{$\propto$}\lower 1.2ex\hbox{$\sim$}}}}
\def\deg{\ifmmode^\circ\else$^\circ$\fi}
\def\pdeg{\ifmmode $\setbox0=\hbox{$^{\circ}$}\rlap{\hskip.11\wd0 .}$^{\circ}
          \else \setbox0=\hbox{$^{\circ}$}\rlap{\hskip.11\wd0 .}$^{\circ}$\fi}
\def\arcs{\ifmmode {^{\scriptstyle\prime\prime}}
          \else $^{\scriptstyle\prime\prime}$\fi}
\def\arcm{\ifmmode {^{\scriptstyle\prime}}
          \else $^{\scriptstyle\prime}$\fi}
\newdimen\sa  \newdimen\sb
\def\parcs{\sa=.07em \sb=.03em
     \ifmmode \hbox{\rlap{.}}^{\scriptstyle\prime\kern -\sb\prime}\hbox{\kern -\sa}
     \else \rlap{.}$^{\scriptstyle\prime\kern -\sb\prime}$\kern -\sa\fi}
\def\parcm{\sa=.08em \sb=.03em
     \ifmmode \hbox{\rlap{.}\kern\sa}^{\scriptstyle\prime}\hbox{\kern-\sb}
     \else \rlap{.}\kern\sa$^{\scriptstyle\prime}$\kern-\sb\fi}
\def\ra[#1 #2 #3.#4]{#1\sup{h}#2\sup{m}#3\sup{s}\llap.#4}
\def\dec[#1 #2 #3.#4]{#1\deg#2\arcm#3\arcs\llap.#4}
\def\deco[#1 #2 #3]{#1\deg#2\arcm#3\arcs}
\def\rra[#1 #2]{#1\sup{h}#2\sup{m}}

\def\dots{\relax\ifmmode \ldots\else $\ldots$\fi}
\def\WHzsr{\ifmmode $W\,Hz\mo\,sr\mo$\else W\,Hz\mo\,sr\mo\fi}
\def\mHz{\ifmmode $\,mHz$\else \,mHz\fi}
\def\GHz{\ifmmode $\,GHz$\else \,GHz\fi}
\def\mKs{\ifmmode $\,mK\,s$^{1/2}\else \,mK\,s$^{1/2}$\fi}
\def\muKs{\ifmmode \,\mu$K\,s$^{1/2}\else \,$\mu$K\,s$^{1/2}$\fi}
\def\muKRJs{\ifmmode \,\mu$K$_{\rm RJ}$\,s$^{1/2}\else \,$\mu$K$_{\rm RJ}$\,s$^{1/2}$\fi}
\def\muKHz{\ifmmode \,\mu$K\,Hz$^{-1/2}\else \,$\mu$K\,Hz$^{-1/2}$\fi}
\def\MJysr{\ifmmode \,$MJy\,sr\mo$\else \,MJy\,sr\mo\fi}
\def\MJysrmK{\ifmmode \,$MJy\,sr\mo$\,mK$_{\rm CMB}\mo\else \,MJy\,sr\mo\,mK$_{\rm CMB}\mo$\fi}
\def\microns{\ifmmode \,\mu$m$\else \,$\mu$m\fi}

\def\muK{\ifmmode \,\mu$K$\else \,$\mu$\hbox{K}\fi}
\def\microK{\ifmmode \,\mu$K$\else \,$\mu$\hbox{K}\fi}
\def\muW{\ifmmode \,\mu$W$\else \,$\mu$\hbox{W}\fi}
\def\kms{\ifmmode $\,km\,s$^{-1}\else \,km\,s$^{-1}$\fi}
\def\kmsMpc{\ifmmode $\,\kms\,Mpc\mo$\else \,\kms\,Mpc\mo\fi}
\providecommand{\sorthelp}[1]{}

\newlength{\arrow}
\newcommand{\planck}[1]{{\textit{Planck}{#1}}}

\def\bda{\boldsymbol{a }}

\def\bdd{\boldsymbol{d }}
\def\bde{\boldsymbol{e }}
\def\bdf{\boldsymbol{f }}
\newcommand{\tA}{\boldsymbol{{\rm A}}}
\newcommand{\tC}{\boldsymbol{{\rm C}}}

\title[Map-based inference of \planck\ SZ temperature]{Evidence for relativistic Sunyaev-Zeldovich effect in \planck\ CMB maps with an average electron-gas temperature of $T_{\rm e}\simeq 5$\,keV}

\author[Remazeilles and Chluba]{
Mathieu Remazeilles$^{1}$\thanks{E-mail: \href{mailto:remazeilles@ifca.unican.es}{remazeilles@ifca.unican.es}} 
and Jens Chluba$^{2}$
\\
$^{1}$Instituto de Física de Cantabria (CSIC-UC), Avda. de los Castros s/n, 39005 Santander, Spain\\
$^{2}$Jodrell Bank Centre for Astrophysics, School of Physics and Astronomy, The University of Manchester, Manchester M13 9PL, UK
}

\date{Accepted XXX. Received YYY; in original form ZZZ}

\pubyear{2025}

\begin{document}
\label{firstpage}
\pagerange{\pageref{firstpage}--\pageref{lastpage}}
\maketitle

\begin{abstract}
Stacking the public \planck\ CMB temperature maps (NILC, SMICA, SEVEM, Commander) on galaxy clusters from \planck\ catalogues reveals substantial residual contamination from thermal Sunyaev-Zeldovich (tSZ) emission. Unexpectedly, stacking "tSZ-free" CMB maps, like the \planck\ SMICA-noSZ or Constrained ILC (CILC) maps, still shows noticeable residual contamination from galaxy clusters. We demonstrate that this persisting residual stems from neglected relativistic SZ (rSZ) corrections in the CMB map estimation. Employing a component-separation method specifically designed for the rSZ effect on \planck\ data, we map the rSZ first-order moment field $y(T_{\rm e}-\bar{T}_{\rm e})$ over the sky for different pivot temperatures $\bar{T}_{\rm e}$ ranging from $2$ to $10$\,keV. Stacking these $y(T_{\rm e}-\bar{T}_{\rm e})$-maps on \planck\ clusters exhibits either an intensity decrement or increment at the centre, contingent upon whether $\bar{T}_{\rm e}$ is above or below the ensemble-averaged cluster temperature $T_{\rm e}$. For the pivot value $\bar{T}_{\rm e}=5$\,keV, a vanishing intensity is observed in the stacked \planck\ $y(T_{\rm e}-\bar{T}_{\rm e})$-map, enabling us to infer the average gas temperature of $T_{\rm e}\simeq 5$\,keV for \planck\ clusters. Building upon this finding, we revisit the \planck\ tSZ-free CMB map by deprojecting the complete rSZ emission using CILC, assuming an rSZ spectrum with $T_{\rm e} = 5$\,keV. Our new, rSZ-free \planck\ CMB map, when stacked on clusters, shows a clear cancellation of residual SZ contamination in contrast to prior (non-relativistic) tSZ-free \planck\ CMB maps. Our map-based approach provides compelling evidence for an average temperature of the \planck\ galaxy clusters of $T_{\rm e} = 4.9 \pm 2.6$\,keV using the rSZ effect.  
\end{abstract}




\section{Introduction}
\label{sec:intro} 

When photons of the cosmic microwave background (CMB) radiation pass through the hot electron gas in galaxy clusters, they are scattered to higher energies, leading to a characteristic spectral distortion of the CMB blackbody emission known as the thermal Sunyaev-Zeldovich (tSZ) effect \citep{Zeldovich1969,Sunyaev1972}. 
In massive clusters, where the electron gas can reach sufficiently high temperatures (typically several keV), the thermal velocities of electrons can approach a significant fraction of the speed of light, making relativistic corrections to the tSZ effect important. These relativistic SZ (rSZ) corrections introduce an additional, subtle spectral distortion that depends on the actual temperature of the electron gas \citep{Rephaeli1995,Challinor1998,Itoh1998, Sazonov1998, Chluba2012}.
From simulation we expect this effect to enter at the level of $\simeq 10\%-20\%$ \citep[e.g.,][]{Kay2008, Lee2022, Kay2024}

Third-generation CMB experiments, like \planck\ \citep{planck2016-l01}, ACT \citep{ACT2020}, and SPT \citep{SPT2011}, lack the sensitivity needed to detect the faint rSZ distortion for individual clusters. Consequently, current SZ data products, including cluster catalogues \citep{planck2014-a36,ACT-SZ-Cat2021,Melin2021,SPT-SZ-Cat2015,Bleem2020} and tSZ Compton $y$-parameter maps \citep{planck2014-a28,Aghanim2019,Madhavacheril2020,Tanimura2022,Bleem2022,Chandran2023,McCarthy2024}, have been based on non-relativistic tSZ formula. 

However, observables that rely on statistically significant samples of clusters, such as the power spectrum of the \planck\ tSZ Compton $y$-map or \planck\ cluster stacking products, have been shown to be sensitive to rSZ temperature corrections \citep{Hurier2016,Erler2018, Remazeilles2019}. Ignoring these corrections in the analysis of current CMB data can lead to non-negligible biases in the estimation of galaxy cluster properties, such as gas pressure profiles \citep{Perrott2024}, and in the determination of cosmological parameters from SZ observables, like the amplitude of dark matter fluctuations, $\sigma_8$ \citep{Remazeilles2019}. As a result, current data are now being revisited to produce relativistic SZ $y$-maps \citep{Remazeilles2019,Coulton2024}.

The tSZ emission from galaxy clusters also acts as a foreground in CMB observations, requiring mitigation through component separation techniques. While this foreground is generally subdominant and mostly invisible locally in CMB maps, stacking the public \planck\ CMB temperature maps -- NILC, SMICA, SEVEM, or Commander \citep{planck2014-a11} -- on galaxy clusters has revealed significant residual contamination from tSZ emission \citep{Chen2018,Madhavacheril2018}. Due to non-Gaussian distribution and correlation with the underlying dark matter distribution, such residual tSZ contamination in CMB maps can affect derived lensing observables and their cross-correlation with other tracers of large-scale structure \citep{vanEngelen2014,Madhavacheril2018,Chen2022}.

The Constrained ILC (CILC) component separation method \citep{Remazeilles2011a} was designed to address this issue by \emph{deprojecting} the tSZ effect from the CMB map to get rid of residual foreground contamination from galaxy clusters. These developments resulted in the creation of noisier but \emph{tSZ-free} CMB maps, such as the \planck\ SMICA-noSZ map \citep{planck2016-l04} and the tSZ-free CMB map from ACT Collaboration \citep{Madhavacheril2020,Coulton2024}. These more constrained CMB maps are supposedly free of residual cluster SZ emission. However, since relativistic corrections to the tSZ formula were not accounted for in the spectral energy distribution (SED) used for deprojection, some cluster residuals still persist even in these tSZ-free CMB maps.
In our study, we search for residual relativistic SZ signals from galaxy clusters within the \planck\ tSZ-free CMB maps. Our aim is to identify potential mismodelling of the thermal SZ effect, such as the omission of relativistic SZ temperature corrections, and to use this information to infer the actual average temperature of \planck\ clusters.

For this endeavour, we propose to stack "tSZ-free" \planck\ CMB maps, where the tSZ emission has been \emph{deprojected} using the CILC technique, on galaxy clusters from the \planck\ cluster catalogue \citep{planck2013-p05a}. If a residual decrement in intensity is observed at the centre of the stacked tSZ-free \planck\ CMB map, it would suggest that the thermal SZ SED used in CILC for the deprojection is not accurately modelled. By incorporating the appropriate relativistic SZ (rSZ) temperature correction into the SED model for the deprojection of the thermal SZ effect in the \planck\ CMB map, and then stacking this revised map on galaxy clusters, we expect to observe a \emph{vanishing} intensity (a null) at the centre.

We also propose to estimate the average temperature of the \planck\ clusters by mapping the first-order moment, $y(T_{\rm e}-\bar{T}_{\rm e})$, of the rSZ effect across the sky using different pivot temperatures $\bar{T}_{\rm e}$ \citep{Remazeilles2020}, and stacking each of these maps on \planck\ galaxy clusters.  The pivot temperature $\bar{T}_{\rm e}=\bar{T}_{\rm e}^{\,*}$ for which the stacked \planck\ $y(T_{\rm e}-\bar{T}_{\rm e}^{\,*})$-map exhibits neither decrement nor increment in intensity but nearly vanishes then indicates the actual average temperature $T_{\rm e}$ of the \planck\ cluster sample.

By incorporating the average cluster temperature $T_{\rm e}=\bar{T}_{\rm e}^{\,*}$ in the SED model for the deprojection of the thermal SZ effect using CILC, we demonstrate how the resulting \planck\ CMB map shows a clear \emph{visual} vanishing of the residual SZ contamination when stacked on \planck\ galaxy clusters, in stark contrast to prior (non-relativistic) tSZ-free \planck\ CMB maps. 
This unique combination of nulling approaches, relying exclusively on the relativistic SZ probe, offers a new original way to infer the average gas temperature of the \planck\ galaxy cluster sample with map-based methods.

This paper is organised as follows. In Section~\ref{sec:data}, we present the \planck\ data used for our analysis. Section~\ref{sec:method} outlines our two complementary methods for detecting the relativistic SZ effect in the \planck\ data and inferring the average temperature of \planck\ clusters. The first method, which focuses on mapping the first-order moment of the relativistic SZ effect and stacking on clusters, is detailed in Section~\ref{sec:szmoment}. The second approach, which involves deprojecting the relativistic SZ effect in CMB maps and stacking on clusters to identify nulls, is described in Section~\ref{sec:cmb}. Our findings are presented in Section~\ref{sec:results}, with results based on \planck\ relativistic SZ moment maps discussed in Section~\ref{subsec:rsz_results}, outcomes from SZ-free \planck\ CMB maps in Section~\ref{subsec:cmb_results}, and semi-analytical scaling relation results in Section~\ref{subsec:scaling_results}. We conclude in Section~\ref{sec:conclusions}.

\section{Data}
\label{sec:data} 

We apply the CILC component separation method to the nine \planck\ frequency maps from the \planck\ release 2 \citep[PR2;][]{planck2014-a01} to generate rSZ-free CMB maps and rSZ first-order moment maps. 
To mitigate residual Galactic foreground contamination in the CILC output maps, we use the \planck\ GAL80 mask, leaving $f_{\rm sky}=80\%$ of the sky unmasked. 
The SED of the relativistic SZ effect for various electron temperatures, which is used in our CILC runs, is modeled with\footnote{{\tt SZpack} is available here: \url{https://github.com/CMBSPEC/SZpack} which also features an improve {\tt python} interface \citep{SZpack2024}.} \textsc{\tt SZpack} \citep{Chluba2012,Chluba2013}. 

The outputs of our CILC runs, whether rSZ-free CMB maps or rSZ first-order moment maps, are stacked on galaxy clusters from the \planck\ PSZ1 catalogue \citep{planck2013-p05a}. We utilize a sample of 829 PSZ1 clusters detected by the \planck's matched multi-frequency filter 1 (MMF1) algorithm, which, after masking the Galactic plane, reduces to 795 clusters. Additionally, we consider a subsample of 327 massive \planck\ clusters with ${M_{500} > 4.5 \times 10^{14}, M_\odot}$ and detected with an $\text{SNR} \geq 4.8$ across the $80\%$ of unmasked sky. 

In addition to the \planck\ CILC CMB maps generated in this work, we also utilise the \planck\ NILC and SMICA-noSZ CMB maps \citep{planck2014-a11}, publicly available in the Planck Legacy Archive,\footnote{\url{https://pla.esac.esa.int/pla}} for comparative analysis.

\section{Nulling approaches for relativistic SZ}
\label{sec:method} 

By \emph{nulling} approaches, we mean searching for a \emph{vanishing} intensity (in contrast to detecting an increment or decrement in intensity) when stacking on galaxy clusters either a CMB map supposedly free from thermal SZ contamination or an rSZ first-order moment map. 

\subsection{\textit{Planck} relativistic SZ first-order moment map stacking}
\label{sec:szmoment} 

The relativistic SZ (rSZ) effect can be expressed as a Taylor expansion around a pivot temperature $\bar{T}_{\rm e}$ as follows \citep{Chluba2013}:
\begin{align}
 \Delta T^{\rm rSZ} &= yf(\nu, T_{\rm e}) \cr 
 			     &\simeq yf(\nu, \bar{T}_{\rm e})\, +\, y\left(T_{\rm e}-\bar{T}_{\rm e}\right)\frac{\partial f(\nu, \bar{T}_{\rm e})}{\partial \bar{T}_{\rm e}}\, +\, \mathcal{O}\left(\left(T_{\rm e}-\bar{T}_{\rm e}\right)^2\right)\,,
\end{align} 
where the Compton parameter $y$ and the temperature $T_{\rm e}$ fields both vary across the sky, but $\bar{T}_{\rm e}$ remains constant, and $f(\nu, T_{\rm e})$ represents the temperature-dependent SED of the rSZ effect. This expansion highlights \emph{moments} of the rSZ temperature field, $y\left(T_{\rm e}-\bar{T}_{\rm e}\right)^k$ where $k\in \mathbb{N}$, that vary across the sky with a uniform frequency dependence (SED), ${\partial^k f(\nu, \bar{T}_{\rm e})/\partial \bar{T}_{\rm e}^k}$, over the sky, enabling effective component separation with CILC.

By employing the component separation method outlined in \cite{Remazeilles2020} on the \planck\ PR2 frequency maps, we can construct first-order moment maps, $y(T_{\rm e}-\bar{T}_{\rm e})$, of the rSZ effect, under different assumptions for the pivot temperature, ranging from $\bar{T}_{\rm e} = 2$ to $10$\,keV, as follows:
\begin{align}
y\left(T_{\rm e}-\bar{T}_{\rm e}\right) = \bde^{\rm T}\left(\tA^{\rm T}\tC^{-1}\tA\right)^{-1}\tA^{\rm T}\tC^{-1}\bdd\,,
\label{eq:momentcilc}
\end{align} 
where $\bdd = \{d_\nu\}$ represents the data vector that compiles the \planck\ multi-frequency maps for all channels, and $\tC = \langle \bdd \bdd^{\rm T} \rangle$ is the empirically estimated frequency-frequency covariance matrix of the data.
The mixing matrix is given as a function of the pivot value $\bar{T}_{\rm e}$ by:
\begin{align}
\label{eq:sedmatrix}
\tA =
\begin{pmatrix}
 \partial \bdf/\partial \bar{T}_{\rm e} & \bdf
\end{pmatrix}\,,
\end{align}
where the first column contains the SED vector of the first-order moment, ${\partial \bdf/\partial \bar{T}_{\rm e} = \{\partial f(\nu, \bar{T}_{\rm e})/\partial \bar{T}_{\rm e}\}}$, evaluated at $T_{\rm e} = \bar{T}_{\rm e}$, and the second column contains the SED vector of the zeroth-order moment, $\bdf = \{f(\nu, \bar{T}_{\rm e})\}$, also evaluated at $T_{\rm e} = \bar{T}_{\rm e}$. The vector 
\begin{align}
\bde^{\rm T} =
\begin{pmatrix}
1 & 0 
\end{pmatrix}
\end{align}
selects the first-order moment while deprojecting the zeroth-order moment.
We then stack each of these $y(T_{\rm e}-\bar{T}_{\rm e})$-maps on galaxy clusters from the \planck\ PSZ1 MMF1 catalogue \citep{planck2013-p05a}. 

The $y(T_{\rm e}-\bar{T}_{\rm e})$-maps are particularly interesting because clusters in these maps appear as intensity increments (red spots) if their actual temperature $T_{\rm e}$ exceeds the chosen pivot temperature $\bar{T}_{\rm e}$, as intensity decrements (blue spots) if $T_{\rm e} < \bar{T}_{\rm e}$, or as nulls if the selected pivot $\bar{T}_{\rm e}$ closely matches to the actual cluster temperature $T_{\rm e}$ \citep{Remazeilles2020}.

By seeking nulls in the stacked \planck\ $y(T_{\rm e}-\bar{T}_{\rm e})$-maps for various pivot values $\bar{T}_{\rm e}$, we can deduce the average temperature of the \planck\ galaxy clusters to be approximately $T_{\rm e}\simeq \bar{T}_{\rm e}^{\,*}$, where $\bar{T}_{\rm e}^{\,*}$ represents the pivot value at which a vanishing intensity is detected during stacking.
This represents the first application of the method described in \cite{Remazeilles2020} on actual \planck\ data.

In addition, relativistic $y$-maps (i.e. zeroth-order moment maps) can also be reconstructed for different pivot temperatures by modifying
\begin{align}
\bde^{\rm T} =
\begin{pmatrix}
1 & 0 
\end{pmatrix}
\qquad \longrightarrow \qquad
\bde^{\rm T} =
\begin{pmatrix}
0 & 1 
\end{pmatrix}
\end{align}
in Eq.~\eqref{eq:momentcilc}. These zeroth-order moment maps can also be stacked on galaxy clusters. This enables us to estimate the average $y$-weighted electron temperature of the \planck\ clusters by calculating the ratio of the first-order moment flux to the zeroth-order moment flux within a central circular aperture $\mathcal{A}$, as described in \cite{Remazeilles2020}:
\begin{align}
\widehat{\Delta T}_{\rm e} = \frac{\langle y(T_{\rm e}-\bar{T}_{\rm e}) \rangle_\mathcal{A}}{\langle y\rangle_\mathcal{A}}
\label{eq:shift}
\end{align} 
and 
\begin{align}
\widehat{T}_{\rm e} = \bar{T}_{\rm e} + \widehat{\Delta T}_{\rm e}\,.
\end{align}
The uncertainties on $\widehat{T}_{\rm e}$ are determined by the mean, variance, and covariance of both the zeroth- and first-order moment maps within the aperture, using the standard formula for the variance of a ratio.

\subsection{\textit{Planck} relativistic SZ-free CMB map stacking}
\label{sec:cmb} 

Building upon the analysis described in Sect~\ref{sec:szmoment}, we can then revisit the construction of a tSZ-free \planck\ CMB temperature map.  This refinement involves integrating the average cluster temperature $\bar{T}_{\rm e}^{\,*}$ into the modelling of the relativistic SZ SED, $f(\nu, \bar{T}_{\rm e}^{\,*})$, using \mbox{\textsc{\tt SZpack}} \citep{Chluba2012,Chluba2013}. This adjusted modelling is utilized to deproject the complete SZ emission out of the CMB map using the CILC component separation method \citep{Remazeilles2011a, Remazeilles2021}. 

This process involves applying the same Eq.~\eqref{eq:momentcilc} to \planck\ data, but with the following modified mixing matrix:
\begin{align}
\label{eq:sedmatrix}
\tA =
\begin{pmatrix}
 \bda & \bdf^*
\end{pmatrix}\,,
\end{align}
where the first column contains the SED vector of the CMB temperature anisotropies, $\bda = \{a(\nu)\}$, which are being reconstructed, and the second column contains the SED vector of the relativistic SZ effect, $\bdf^* = \{f(\nu, \bar{T}^*_{\rm e})\}$, evaluated at $T_{\rm e} = \bar{T}^*_{\rm e}$, which is being deprojected.
The resulting rSZ-deprojected \planck\ CMB map (assuming $T_{\rm e} =  \bar{T}_{\rm e}^{\,*}$) can then be stacked on galaxy clusters and compared with prior tSZ-deprojected \planck\ CMB maps from CILC and SMICA-noSZ \citep{planck2016-l04} (which implicitly assumed non-relativistic tSZ) to provide further \emph{visual} map-based evidence for the actual average temperature of the \planck\ galaxy clusters.

\subsection{Needlet-based implementation}
\label{sec:needlets} 

The covariance matrix and CILC weights in Eq.~\eqref{eq:momentcilc} are computed in a needlet basis by decomposing each frequency map into needlet coefficient maps, as described in the section~2.4 of \cite{Remazeilles2020}. In this work, the decomposition is performed over ten needlet windows in harmonic space, as defined in the equation~7 and figure~3 of \cite{Chandran2023}.

For each needlet window, the covariance matrix is computed at each pixel for all frequency pairs by averaging the product of the two frequency (needlet) maps over a pixel domain centred on the target pixel. In practice, this is achieved by smoothing the product of the two frequency (needlet) maps with a Gaussian beam, whose FWHM is determined by the characteristic angular scale selected by the corresponding needlet window. Therefore, a covariance matrix map is obtained for each needlet scale, which in turn provides CILC weight maps for each needlet scale and frequency.

These weight maps are subsequently applied to the needlet coefficient maps of each frequency channel. For the first three needlet windows (corresponding to low multipoles), all nine \planck\ channels are used in CILC. However, for the remaining seven needlet windows, only the six \planck\ HFI channels are included, as the three \planck\ LFI channels lack the necessary angular resolution.

\section{Average temperature of \textit{Planck} clusters}
\label{sec:results} 

In this section, we present our results regarding the estimation of the average electron gas temperature of the  \planck\ galaxy clusters, using the two map-based approaches introduced in Section~\ref{sec:method}.

\subsection{Estimation from \planck\ rSZ first-order moment maps}
\label{subsec:rsz_results}

\begin{figure}
    \centering
    \includegraphics[width=0.8\columnwidth]{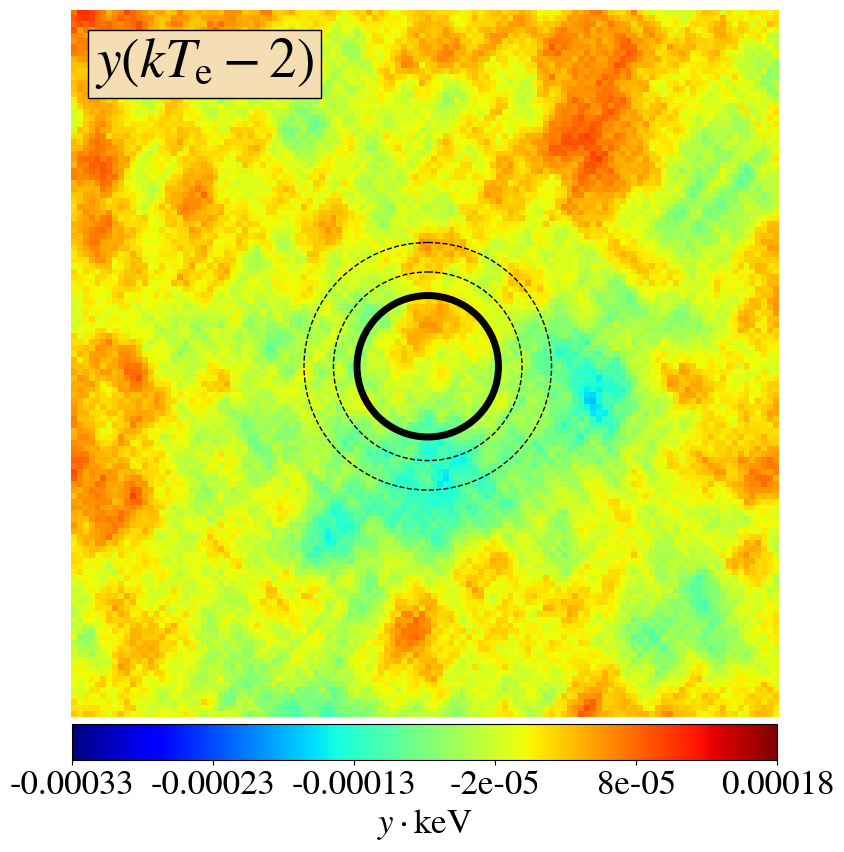}\\
    \includegraphics[width=0.8\columnwidth]{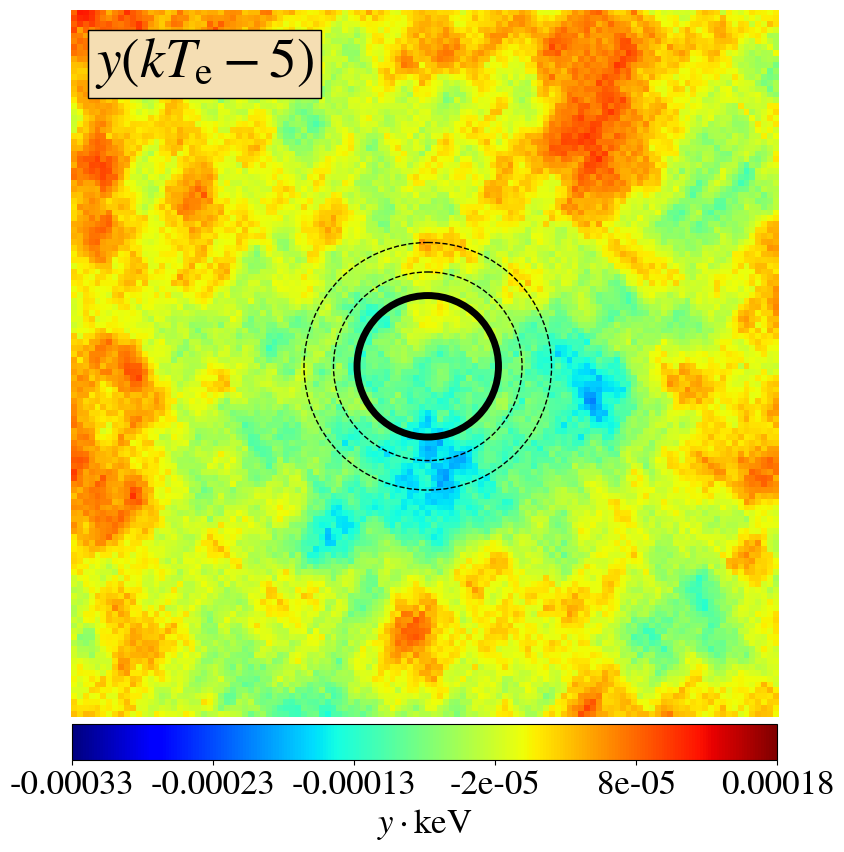}\\
    \includegraphics[width=0.8\columnwidth]{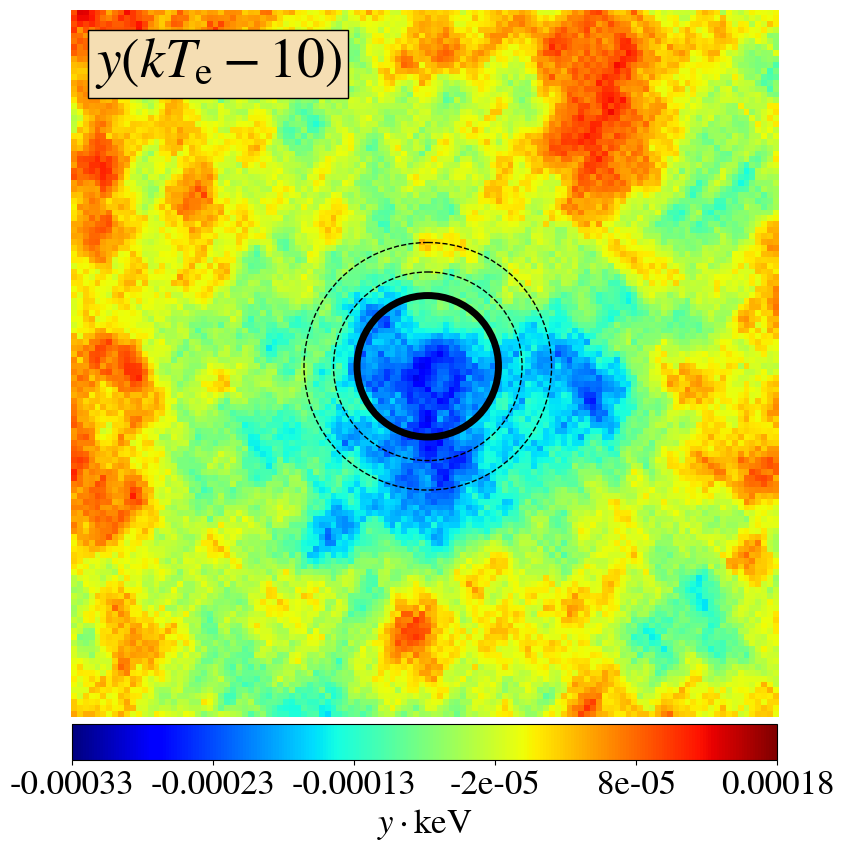}
    \caption{\planck\ rSZ first-order moment maps, $y(T_{\rm e}-\bar{T}_{\rm e})$, stacked on all 795 galaxy clusters of the \planck\ PSZ1 MMF1 catalogue,  for pivot temperatures $\bar{T}_{\rm e}=2$\,keV (\emph{top}), $\bar{T}_{\rm e}=5$\,keV (\emph{middle}) and  $\bar{T}_{\rm e}=10$\,keV  (\emph{bottom}). The dimensions of the patches are $45'\times 45'$, while the outlined circle has a radius of $4.5'$. The stacked $y(T_{\rm e}-\bar{T}_{\rm e})$-map exhibits an increment in intensity (red spot) around the centre for $\bar{T}_{\rm e}=2$\,keV (\emph{top}), a null for $\bar{T}_{\rm e}=5$\,keV (\emph{middle}), and a decrement in intensity (blue spot) for $\bar{T}_{\rm e}=10$\,keV (\emph{bottom}).}
    \label{fig:moment}
\end{figure}

\begin{figure}
    \centering
    \includegraphics[width=0.8\columnwidth]{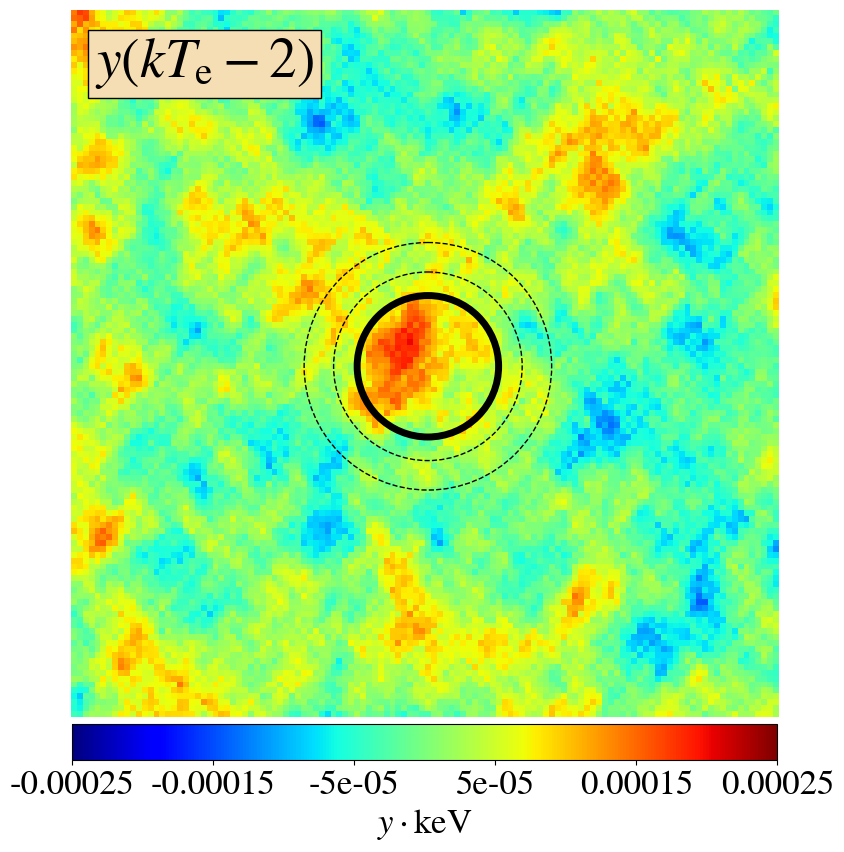}\\
    \includegraphics[width=0.8\columnwidth]{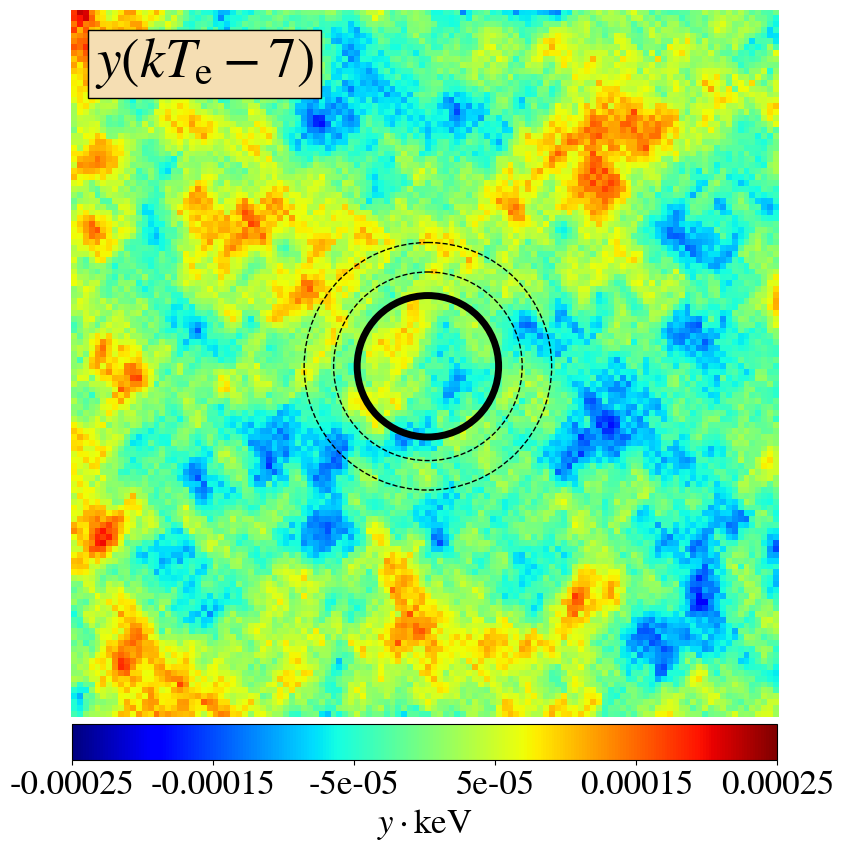}\\
    \includegraphics[width=0.8\columnwidth]{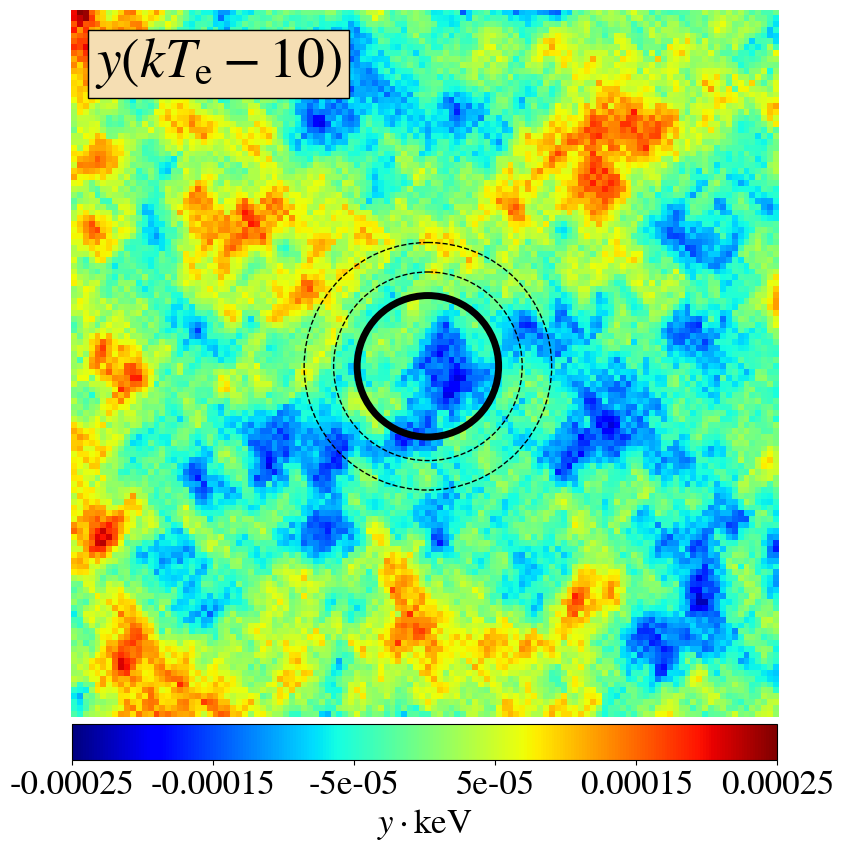}
    \caption{Same as Figure~\ref{fig:moment} but after stacking on the 327 most massive galaxy clusters of the \planck\ PSZ1 MMF1 catalogue with ${M_{500} > 4.5 \times 10^{14}\, M_\odot}$ and detected with $\text{SNR} \geq 4.8$, for pivot temperatures $\bar{T}_{\rm e}=2$\,keV (\emph{top}), $\bar{T}_{\rm e}=7$\,keV (\emph{middle}) and $\bar{T}_{\rm e}=10$\,keV (\emph{bottom}). For this sub-sample of massive \planck\ clusters, the stacked map exhibits an increment in intensity (red spot) around the centre for $\bar{T}_{\rm e}=2$\,keV (\emph{top}), a null for $\bar{T}_{\rm e}=7$\,keV (\emph{middle}), and a decrement in intensity (blue spot) for $\bar{T}_{\rm e}=10$\,keV (\emph{bottom}).}
    \label{fig:moment_masscut}
\end{figure}

Figure~\ref{fig:moment} shows the \planck\ rSZ first-order moment maps, $y(T_{\rm e}-\bar{T}_{\rm e})$, for pivot temperatures $\bar{T}_{\rm e}=2$\,keV, $5$\,keV and $10$\,keV, stacked on 795 galaxy clusters from the \planck\ PSZ1 MMF1 cluster catalogue. Each map has an angular resolution of $5'$. While the residual background RMS contamination outside the circular aperture monotonically increases with rising $\bar{T}_{\rm e}$ (from top to bottom panel), the intensity within the central aperture exhibits a significantly different behaviour. An increment of intensity is noticeable around the centre of the patch for $\bar{T}_{\rm e}=2$\,keV. In contrast, a null intensity is detected for $\bar{T}_{\rm e}=5$\,keV, while a significant decrement of intensity is evident for $\bar{T}_{\rm e}=10$\,keV. Based on these outcomes, we infer that the average temperature $T_{\rm e}$ of the \planck\ clusters must be higher than $\bar{T}_{\rm e}=2$\,keV, lower than $\bar{T}_{\rm e}=10$\,keV, and reasonably close to $\bar{T}_{\rm e}=5$\,keV. Note the offset colour scale towards negative values, intended to highlight the peculiar behaviour of the central cluster signal, as it sits on top of a large-scale negative residual (much beyond the cluster size). High-pass filtering of the stacked $y(T_{\rm e}-\bar{T}_{\rm e})$ maps eliminates this large-scale residual, as discussed in Appendix~\ref{sec:app1}.

Figure~\ref{fig:moment_masscut} presents a similar stacking analysis of the $y(T_{\rm e}-\bar{T}_{\rm e})$-maps, but focuses on the 327 most massive \planck\ PSZ1 MMF1 galaxy clusters with ${M_{500} > 4.5 \times 10^{14}, M_\odot}$ and detected with $\text{SNR} \geq 4.8$ across the $80\%$ of unmasked sky. In this case, an  increment of intensity is observed for pivot values $\bar{T}_{\rm e} < 7$\,keV, a decrement for pivot values $\bar{T}_{\rm e} > 7$\,keV, and a null intensity when $\bar{T}_{\rm e}=7$\,keV, highlighting that massive \planck\ clusters have a higher gas temperature on average. This trend is consistent with mass-temperature scaling relations \citep{Arnaud2005,Lee2020,Lee2022}.

Table~\ref{tab:moment_flux} displays the integrated fluxes of the stacked ${y(T_{\rm e}-\bar{T}_{\rm e})}$-maps for pivot temperatures ranging from $\bar{T}_{\rm e}=2$ to $10$\,keV, determined through aperture photometry within a $4.5'$ circular aperture (depicted as a solid black circle in Fig.~\ref{fig:moment}). Background flux correction was applied using an annulus with an inner radius of $6'$ and an outer radius of $8'$ (outlined by the dotted black circles in Fig.~\ref{fig:moment}). The second column of Table~\ref{tab:moment_flux} shows the results for stacking all \planck\ PSZ1 MMF1 clusters, while the third column provides the results for the most massive \planck\ PSZ1 MMF1 clusters. As $\bar{T}_{\rm e}$ increases from $2$ to $10$\,keV, the integrated flux transitions from positive to negative values, reaching its minimum absolute value at ${\bar{T}_{\rm e}=5}$\,keV when stacking on all clusters and at ${\bar{T}_{\rm e}=7}$\,keV when stacking on the most massive clusters. This suggests that the average temperature of all detected \planck\ clusters is about $T_{\rm e}\simeq 5$\,keV, while for the most massive clusters, it is around $T_{\rm e}\simeq 7$\,keV. 

    \begin{table}
    \centering
     \caption{Integrated flux in $\text{keV}\cdot\text{arcmin}^2$ units of the stacked \planck\ ${y(T_{\rm e}-\bar{T}_{\rm e})}$-maps stacked on PSZ1 MMF1 clusters within the central circular aperture of radius $4.5'$ (outlined circle in Fig.~\ref{fig:moment}), corrected for the background flux within the annulus of inner radius $6'$ and outer radius $8'$. The integrated flux varies from positive to negative values as $\bar{T}_{\rm e}$ increases, reaching a minimum absolute value for $\bar{T}_{\rm e} = 5$\,keV when stacking on all 795 PSZ1 MMF1 clusters outside the Galactic mask and $\bar{T}_{\rm e} = 7$\,keV when stacking on the 327 most massive clusters with $M_{500} > 4.5 \times 10^{14}\, M_\odot$ (and $\text{SNR} > 4.8$) only.} 
     \label{tab:moment_flux}
     \begin{tabular}{ccc}
        \hline
      	 SZ moment map	& Flux & Flux \\
        $y(T_{\rm e}-\bar{T}_{\rm e})$   &  (all clusters) &   (massive clusters) \\
        \hline
          $y(T_{\rm e}-2)$ & $0.0017 \pm 0.00020$ & $0.0040 \pm 0.00017$  \\ 
          $y(T_{\rm e}-3)$ & $0.0012 \pm 0.00021$ &  $0.0033 \pm 0.00018$ \\
          $y(T_{\rm e}-4)$ & $0.0006 \pm 0.00022$ & $0.0026 \pm 0.00019$ \\
          $y(T_{\rm e}-5)$ & $-0.0001 \pm 0.00023$ & $0.0018  \pm 0.00020$  \\
          $y(T_{\rm e}-6)$ & $-0.0008 \pm 0.00023$ & $0.0011 \pm 0.00021$  \\
          $y(T_{\rm e}-7)$ & $-0.0015 \pm 0.00024$ & $0.0003 \pm 0.00022$  \\
          $y(T_{\rm e}-8)$ & $-0.0022 \pm 0.00025$ & $-0.0006 \pm 0.00023$  \\
          $y(T_{\rm e}-9)$ & $-0.0030 \pm 0.00026$ & $-0.0015 \pm 0.00025$  \\
          $y(T_{\rm e}-10)$ & $-0.0038 \pm 0.00027$ & $-0.0025 \pm 0.00026$ \\
        \hline
     \end{tabular}
    \end{table}
    
    \begin{figure}
    \centering
    \includegraphics[width=\columnwidth]{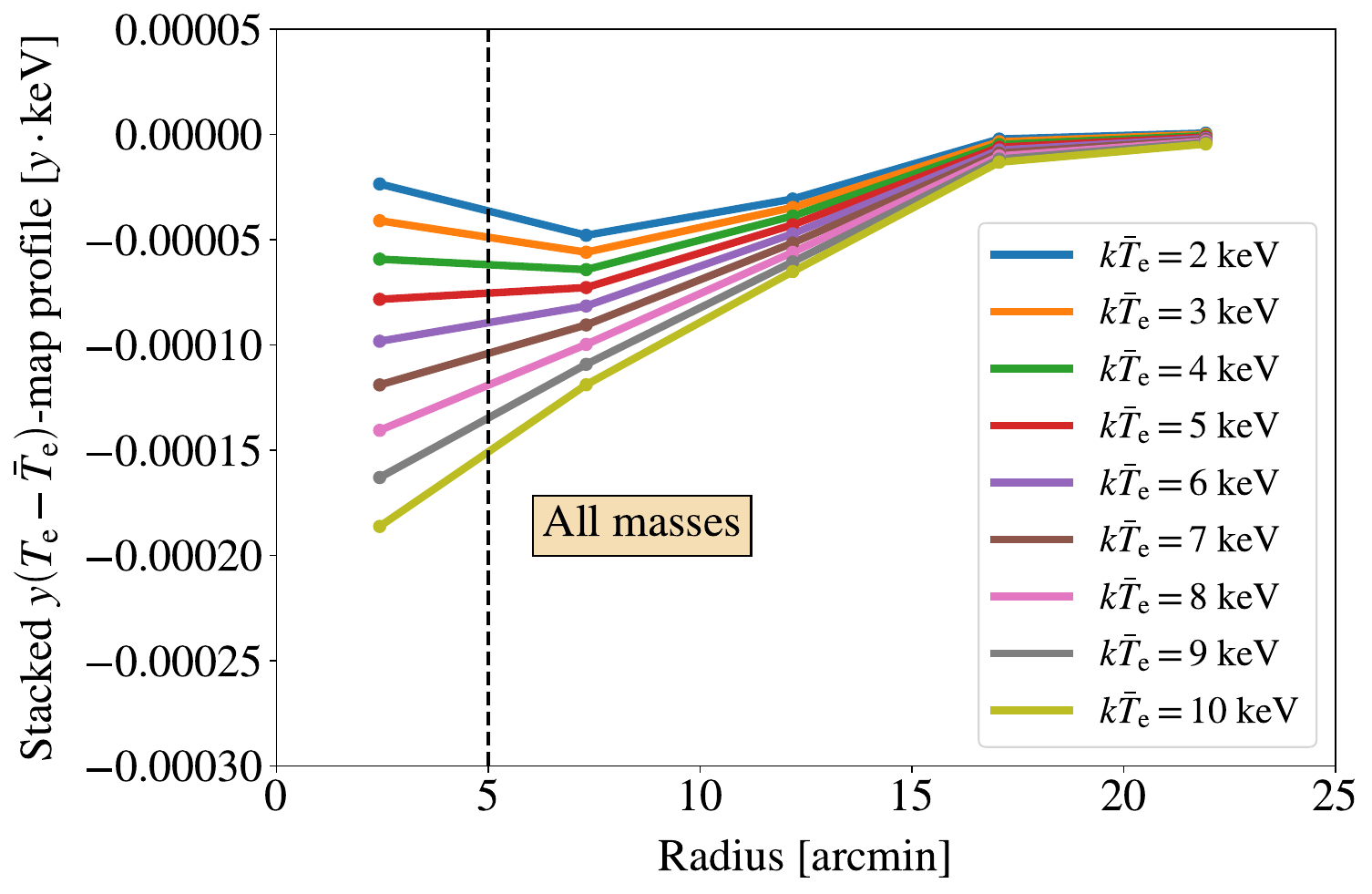}\\
     \includegraphics[width=\columnwidth]{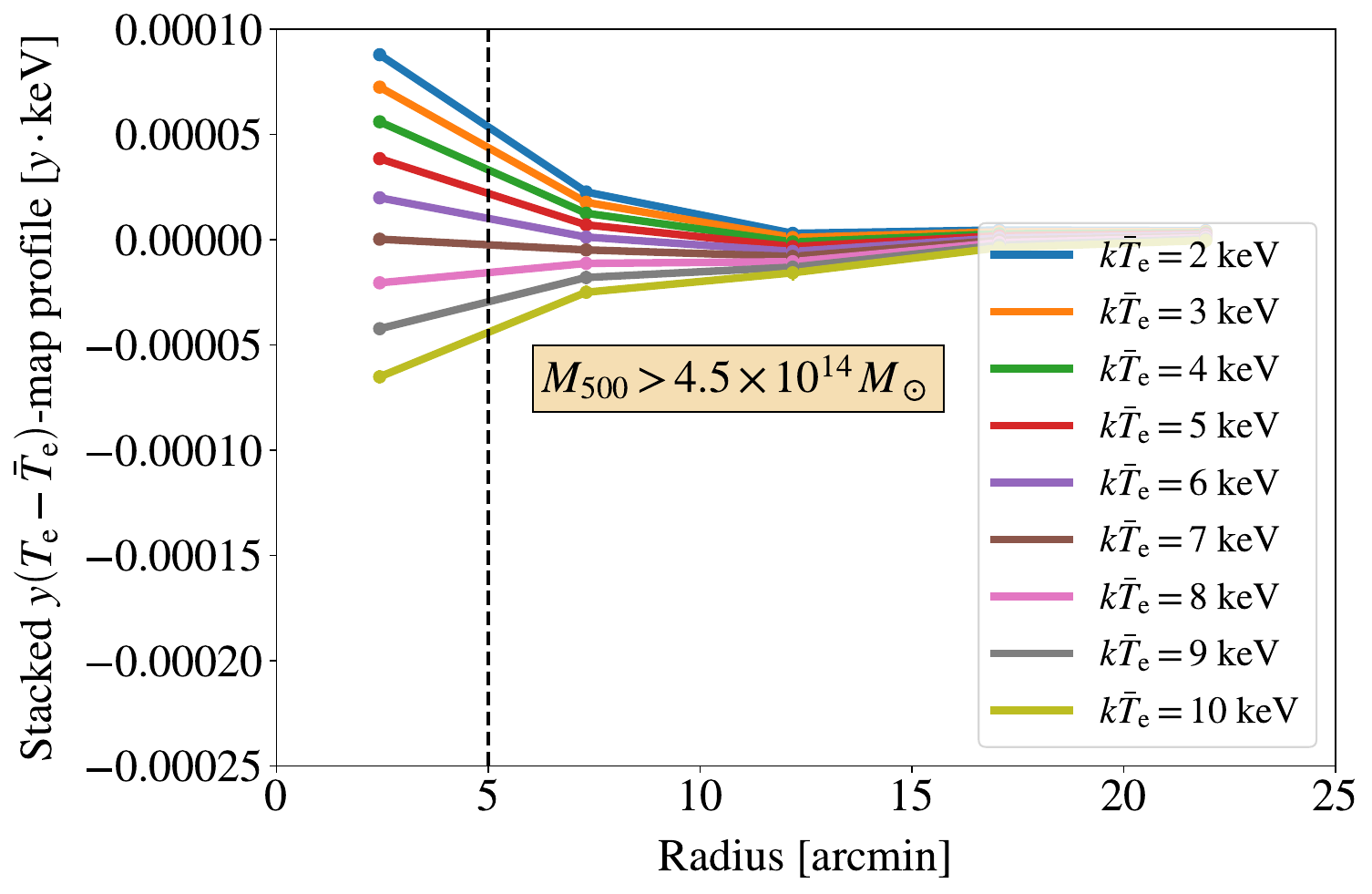}~
    \caption{Profiles of the \planck\ $y(T_{\rm e}-\bar{T}_{\rm e})$-maps stacked on \planck\ PSZ1 MMF1 galaxy clusters (Figures.~\ref{fig:moment} and \ref{fig:moment_masscut})  for various pivot temperatures $\bar{T}_{\rm e}$. Within a $5'$ radius outlined by the vertical dashed line, the slope of the profile of the stacked \planck\ rSZ moment map is positive towards the cluster centre for $k\bar{T}_{\rm e}=2$\,keV (\emph{blue}), flat for $k\bar{T}_{\rm e}=5$\,keV (\emph{red}) when stacking on all clusters (\emph{top}) or $k\bar{T}_{\rm e}=7$\,keV (\emph{brown}) when stacking on most massive clusters (\emph{bottom}), and negative for $\bar{T}_{\rm e}=10$\,keV (\emph{olive}). The negative offset in the top panel is due to the stacked cluster signal sitting on a large-scale (much beyond the cluster size) negative residual, but the key point is the change in the profile's slope towards the centre (see Appendix~\ref{sec:app1} for further discussion).}
    \label{fig:moment_profile}
\end{figure}

      \begin{table}
    \centering
     \caption{Ratio of the first- to zeroth-moment rSZ fluxes, ${\widehat{\Delta T}_{\rm e} = \frac{\langle y(T_{\rm e}-\bar{T}_{\rm e}) \rangle_\mathcal{A}}{\langle y\rangle_\mathcal{A}}}$, within the circular aperture $\mathcal{A}$, along with  the associated uncertainty, for different pivot temperatures $\bar{T}_{\rm e}$. The estimated deviation $\widehat{\Delta T}_{\rm e}$ relative to the assumed pivot  temperature $\bar{T}_{\rm e}$ transitions from positive to negative values as $\bar{T}_{\rm e}$ increases, reaching a minimum absolute value at $\bar{T}_{\rm e} = 5$\,keV for the full cluster sample, and $\bar{T}_{\rm e} = 7$\,keV for the subsample of the most massive clusters. The resulting estimate, $\widehat{T}_{\rm e} = \bar{T}_{\rm e}+\widehat{\Delta T}_{\rm e}$,  of the average cluster temperature is $\widehat{T}_{\rm e} = (4.9\pm 2.6)$\,keV for the full cluster sample, and $\widehat{T}_{\rm e} = (7.3\pm 2.1)$\,keV for the subsample of the most massive clusters.} 
     \label{tab:temperature_deviation}
     \begin{tabular}{ccc}
        \hline
      	 Pivot $\bar{T}_{\rm e}$ [keV] 	& $\widehat{\Delta T}_{\rm e}$ [keV]  & $\widehat{\Delta T}_{\rm e}$ [keV] \\
          &  (all clusters) &   (massive clusters) \\
        \hline
          $2$ & $2.54  \pm  2.22$ & $4.19 \pm 1.70$  \\ 
          $3$ & $1.68 \pm 2.36$ &  $3.45\pm 1.78$ \\
          $4$ & $0.83 \pm 2.50$ &  $2.67 \pm 1.86$ \\
          $5$ & $-0.15 \pm 2.65$ &  $1.88 \pm 1.94$ \\
          $6$ & $-1.17 \pm 2.80$ &  $1.11 \pm 2.03$ \\
          $7$ & $-2.15 \pm 2.96$ &  $0.26 \pm 2.12$ \\
          $8$ & $-3.24 \pm 3.12$ &  $-0.58 \pm 2.21$ \\
          $9$ & $-4.34 \pm 3.29$ &  $-1.57 \pm 2.30$ \\
          $10$ & $-5.53 \pm 3.46$ &  $-2.58 \pm 2.39$ \\
          \hline
     \end{tabular}
    \end{table}

  \begin{figure}
    \centering
    \includegraphics[width=\columnwidth]{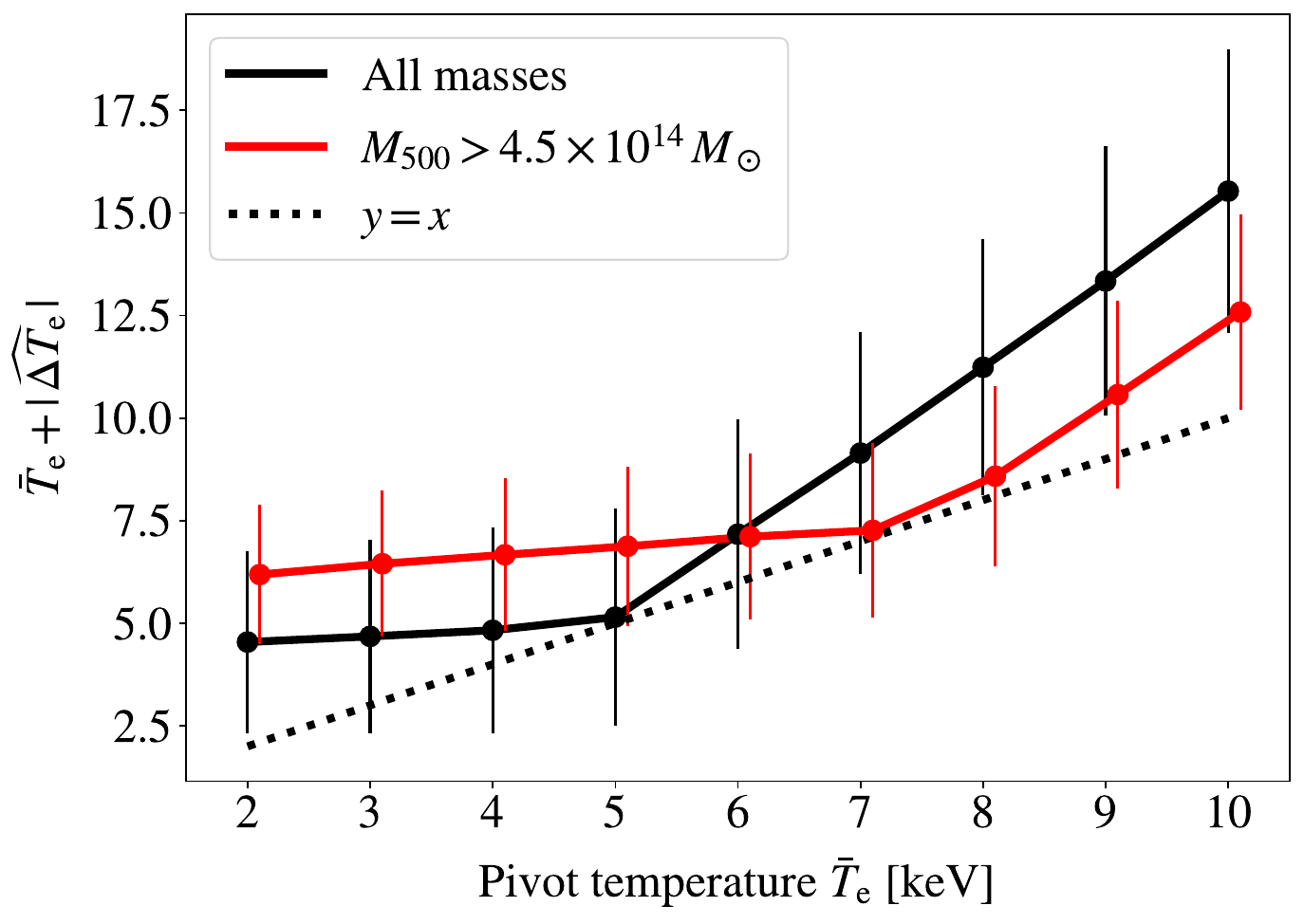}~
     \caption{Absolute deviations, $\bar{T}_{\rm e} + \vert\widehat{\Delta T}_{\rm e}\vert$, around the assumed pivot temperatures $\bar{T}_{\rm e}$, along with the associated uncertainties on the temperature estimates $\widehat{T}_{\rm e} = \bar{T}_{\rm e}+\widehat{\Delta T}_{\rm e}$. The minimum absolute deviation is obtained for $\bar{T}_{\rm e}=5$\,keV, resulting in $\widehat{T}_{\rm e} = (4.9\pm 2.6)$\,keV for the full cluster sample (black), and for $\bar{T}_{\rm e}=7$\,keV, giving $\widehat{T}_{\rm e} = (7.3\pm 2.1)$\,keV for the subsample of the most massive clusters (red).}
    \label{fig:u-shapes}
\end{figure}

Figure~\ref{fig:moment_profile} shows the profiles of the stacked \planck\ $y(T_{\rm e}-\bar{T}_{\rm e})$-maps for different pivot temperatures and different cluster mass samples. These profiles were computed by integrating the signal within consecutive rings of $5'$ width (equivalent to 13 pixels for a pixel size of $0.375'$). Consistently with Figs.~\ref{fig:moment}-\ref{fig:moment_masscut}, within a $5'$ radius, the slope of the stacked \planck\ $y(T_{\rm e}-\bar{T}_{\rm e})$ profile increases toward the centre for $\bar{T}_{\rm e}=2$\,keV, decreases toward the centre for $\bar{T}_{\rm e}=10$\,keV, and remains relatively flat for intermediate pivot temperatures like $\bar{T}_{\rm e}=5$\,keV for the full cluster sample and $\bar{T}_{\rm e}=7$\,keV for the most massive cluster sample. These findings further support the evidence for an average temperature of $T_{\rm e}\simeq 5$\,keV for the \planck\ clusters.
 
Table~\ref{tab:temperature_deviation} presents the effective deviation, $\widehat{\Delta T}_{\rm e}$, of the average cluster temperature from the assumed pivot temperature, $\bar{T}_{\rm e}$, for various pivot values. This deviation is estimated from the ratio of the first-order moment flux to the zeroth-order moment flux (Eq.~\ref{eq:shift}). As $\bar{T}_{\rm e}$ increases, the temperature deviation transitions from positive to negative values, reaching its smallest absolute value at $\bar{T}_{\rm e} = 5$\,keV for the full cluster sample, and at $\bar{T}_{\rm e} = 7$\,keV for the subsample of the most massive clusters. The absolute deviation from different pivot temperatures is also shown in Fig.~\ref{fig:u-shapes}, which highlights a minimal deviation around $\bar{T}^*_{\rm e} = 5$\,keV (black line) or $\bar{T}^*_{\rm e} = 7$\,keV (red line) depending on the cluster sample.

These results allow us to estimate the average cluster temperature as ${\widehat{T}_{\rm e} = \bar{T}^*_{\rm e} + \widehat{\Delta T}_{\rm e}}$, yielding $\widehat{T}_{\rm e} = (4.9 \pm 2.6)$\,keV for the full sample and $\widehat{T}_{\rm e} = (7.3 \pm 2.1)$\,keV for the subsample of the most massive clusters.\footnote{In contrast, stacking on the complementary subsample of the least massive \planck\ clusters produced results with very low statistical significance, yielding $\widehat{T}_{\rm e} = (2.5 \pm 4.2)$\,keV.} Our map-based blind analysis estimates of the \planck\ rSZ temperature are broadly consistent with the findings of \cite{Erler2018}, who reported $T=4.4^{+2.1}_{-2.0}$\,keV and $T=6.0^{+3.8}_{-2.9}$\,keV using parametric SED fitting.

\subsection{Estimation from SZ-free \planck\ CMB maps}
\label{subsec:cmb_results}

\begin{figure}
    \centering
    \includegraphics[width=0.73\columnwidth]{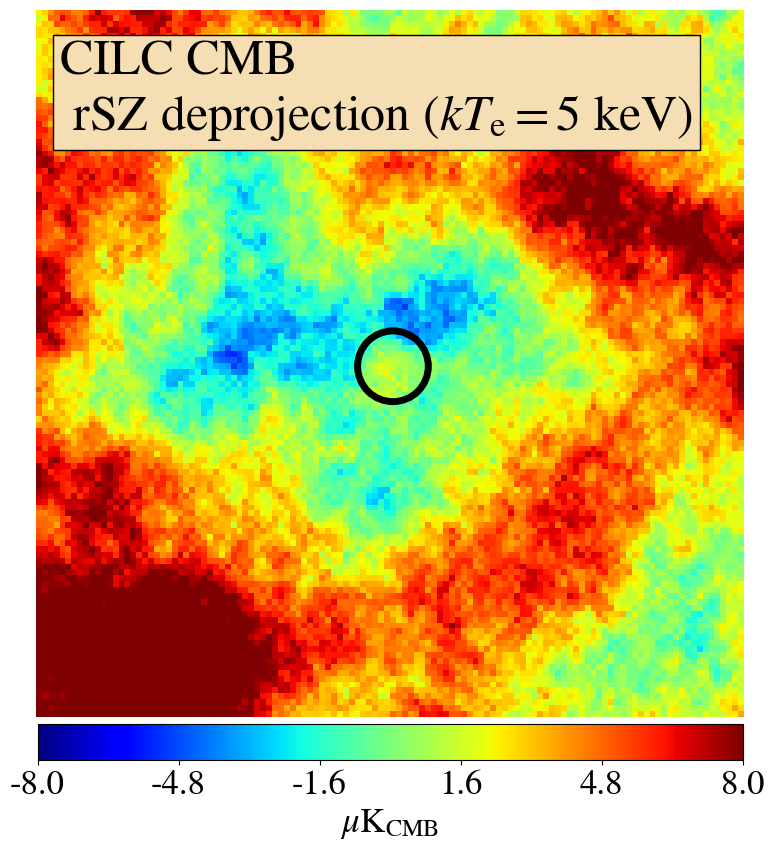}\\
    \includegraphics[width=0.73\columnwidth]{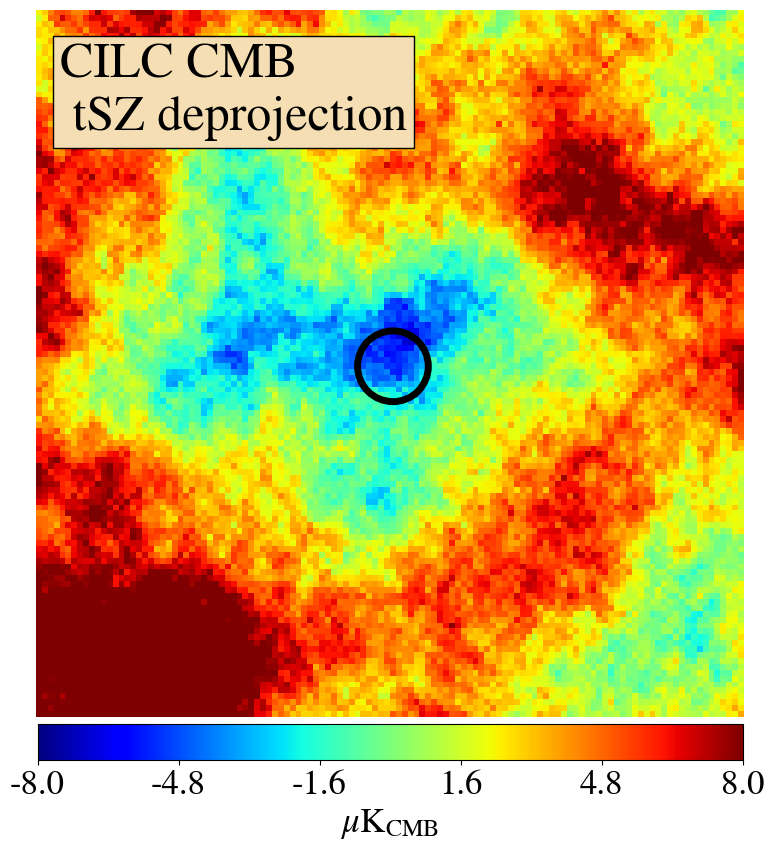}\\
    \includegraphics[width=0.73\columnwidth]{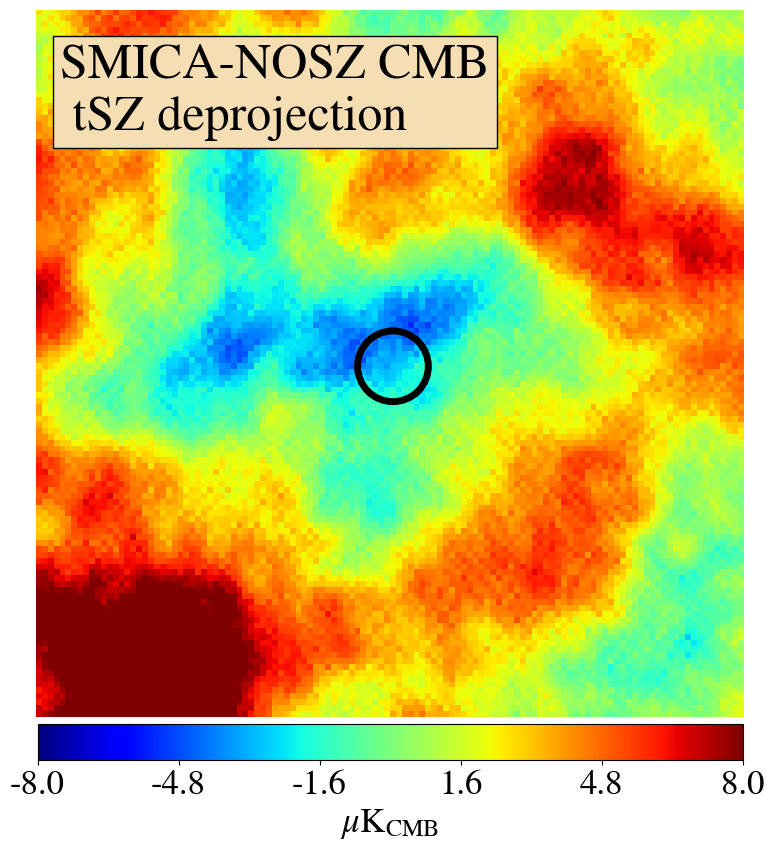}
        \caption{\planck\ CMB maps stacked on 795 \planck\ clusters, after deprojection of the tSZ effect with different temperature assumptions. \emph{Top}:  rSZ-deprojected CILC CMB map, assuming $T_{\rm e}=5$\,keV.  \emph{Middle}:  tSZ-deprojected CILC CMB map, implicitly assuming $T_{\rm e}=0$. \emph{Bottom}:  tSZ-deprojected SMICA-noSZ CMB map, implicitly assuming $T_{\rm e}=0$. Each patch measures $1.5^\circ\times 1.5^\circ$. The two non-relativistic tSZ-free CMB maps show a noticeable decrement (blue spot) at the centre due to residual contamination from galaxy clusters, which vanishes in the relativistic SZ-free CMB map (\emph{top}) by using $T_{\rm e}=5$\,keV in the deprojection model. The small zero-level difference between the SMICA and CILC maps has been removed for comparison.}
    \label{fig:cmb}
\end{figure}

Figure~\ref{fig:cmb} shows the stacking of \planck\ CMB maps on galaxy clusters after deprojection of the thermal SZ effect, using different assumptions for the thermal SZ SED. The angular  resolution of these maps is $5'$. The \planck\ CILC CMB map and the public \planck\ SMICA-noSZ CMB map, for which tSZ has been deprojected using the \emph{non-relativistic} limit of the tSZ SED, are displayed in the middle and bottom panels, respectively. Although the SMICA map has the same angular resolution as the CILC map, it appears smoother due to its lower noise level.
Despite the expectation of being free from residual tSZ contamination by clusters, both cases exhibit a noticeable residual decrement (blue spot) at the centre of the stacked patches.

In contrast, the top panel shows the stacked \planck\ CILC CMB map, where the full relativistic SZ SED is employed instead of the non-relativistic SED, assuming an average electron gas temperature of ${T_{\rm e} = 5}$\,keV, as indicated by our previous findings. In this case, a distinct cancellation of the tSZ residual is evident at the centre of the stacked CMB map within a $5'$ radius. This cancellation starkly contrasts with the persisting residual observed when stacking conventional tSZ-free \planck\ CMB maps (middle and bottom panels). It provides compelling visual evidence for the relativistic SZ effect in the \textit{Planck} data and also supports an average temperature of $T_{\rm e}\simeq 5$\,keV for the \textit{Planck} galaxy clusters. 

\begin{figure}
    \centering
    \includegraphics[width=0.73\columnwidth]{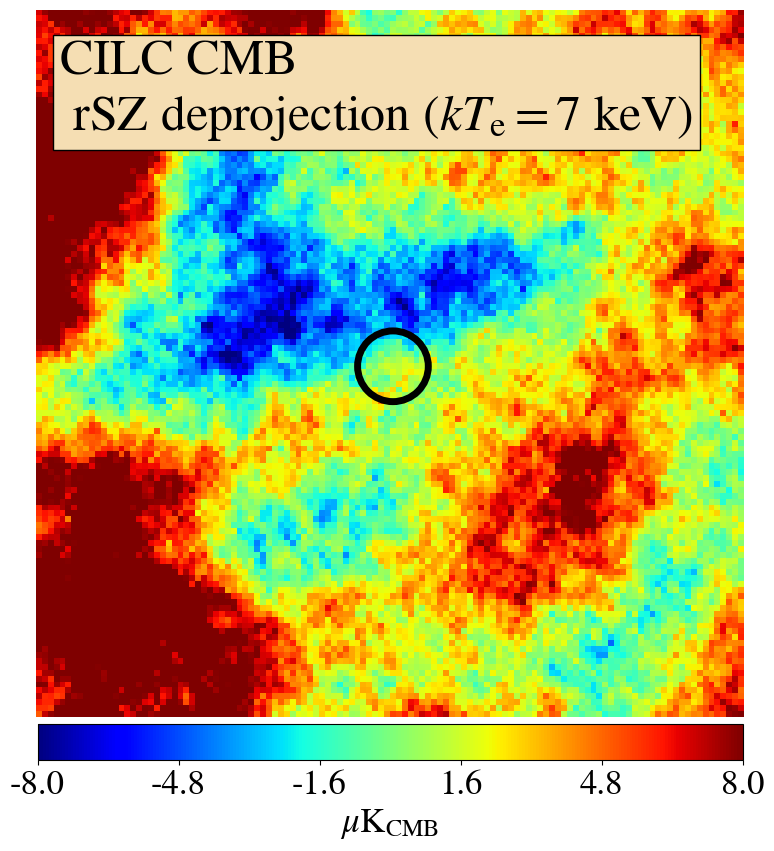}\\
    \includegraphics[width=0.73\columnwidth]{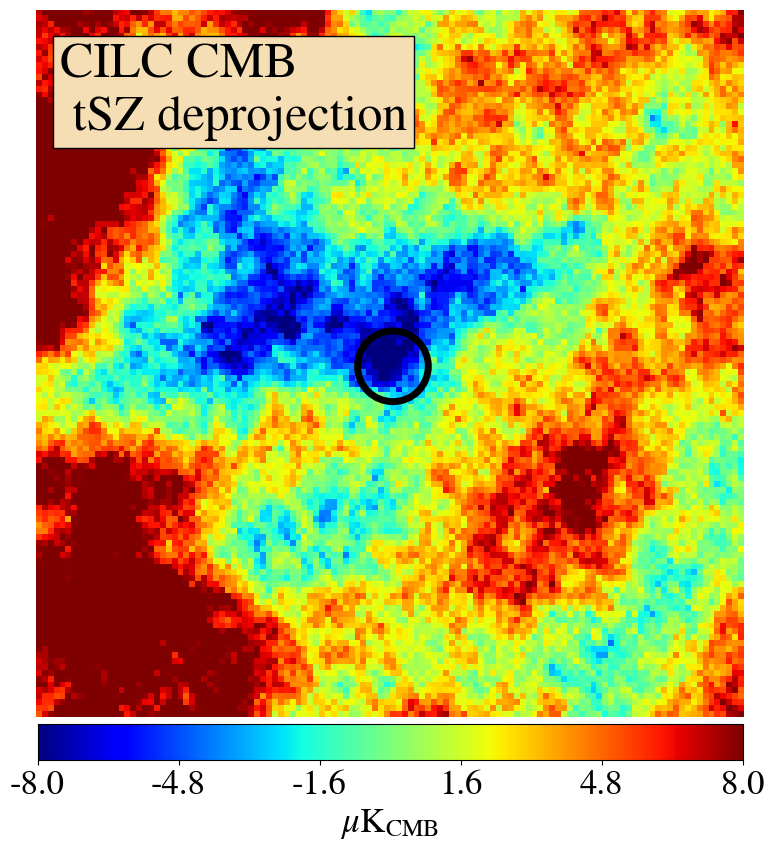}\\
    \includegraphics[width=0.73\columnwidth]{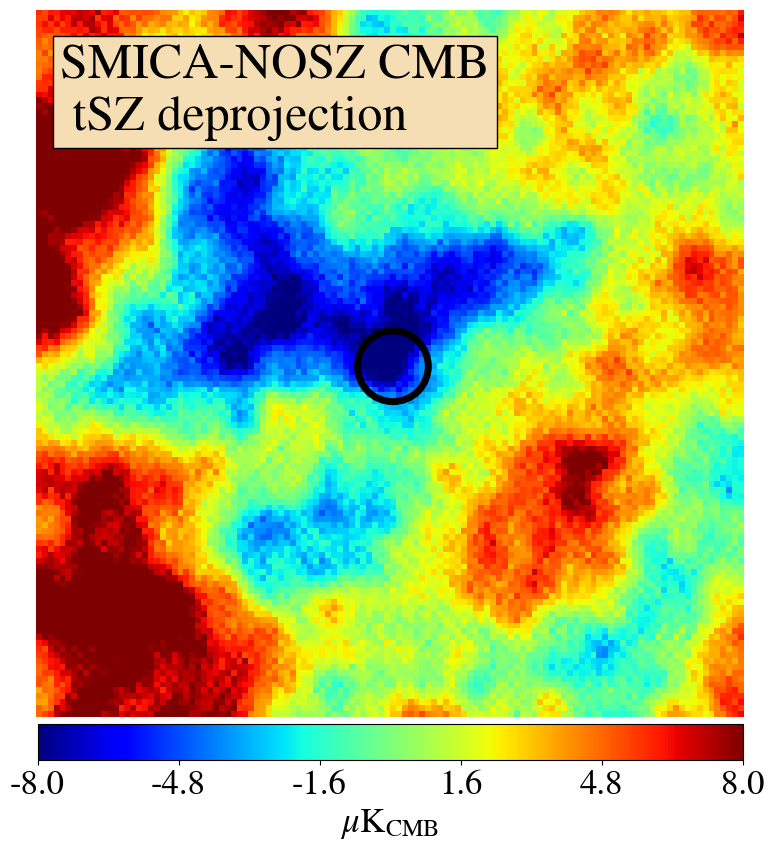}
    \caption{Same as Figure~\ref{fig:cmb} but after stacking on the 372 most massive \planck\ clusters with ${M_{500} > 4.5 \times 10^{14}\, M_\odot}$. In this case, the residual contamination from galaxy clusters effectively vanishes in the rSZ-free \planck\ CMB map (\emph{top}) when using the average cluster temperature $T_{\rm e}=7$\,keV in the SED model for the deprojection.}
    \label{fig:cmb_masscut}
\end{figure}

Figure~\ref{fig:cmb_masscut} shows similar findings after stacking on the 372 most massive \planck\ clusters in the sample, with ${M_{500} > 4.5 \times 10^{14}\, M_\odot}$. In this case, the top panel demonstrates a clear cancellation of the cluster residuals in the \planck\ CILC CMB map when the relativistic SZ effect is deprojected, assuming an average electron temperature of ${T_{\rm e} = 7}$\,keV. In contrast, the middle and bottom panels, which display the conventional tSZ-free \planck\ CMB maps, still exhibit a significant residual decrement (blue spot) at the centre due to ignoring relativistic SZ temperature corrections.

Table~\ref{tab:cmb_flux} displays the integrated flux of several stacked CMB maps within a $4.5'$ circular aperture (outlined by a black circle in Fig.~\ref{fig:cmb}), normalized by the aperture area. The public \planck\ NILC CMB map shows the most significant negative flux, attributed to residual tSZ contamination, as no deprojection was applied in this case. The tSZ-free \planck\ CILC and SMICA-noSZ CMB maps exhibit substantially lower fluxes compared to the \planck\ NILC CMB map due to the deprojection of the non-relativistic thermal SZ effect. The rSZ-free \planck\ CILC CMB maps, where relativistic temperature corrections have been incorporated into the deprojection, display the lowest flux values, indicating effective nulling of residual SZ contamination. As the assumed temperature increases, the residual flux in the stacked rSZ-free CMB maps shifts from negative to positive values, suggesting slight under- or overestimation of the average cluster temperature. The minimum flux is observed when deprojecting the full rSZ emission at $T_{\rm e} = 5$\,keV for the entire cluster sample and at $T_{\rm e} = 7$\,keV for the subsample of massive clusters.

\begin{figure}
    \centering
     \includegraphics[width=0.85\columnwidth]{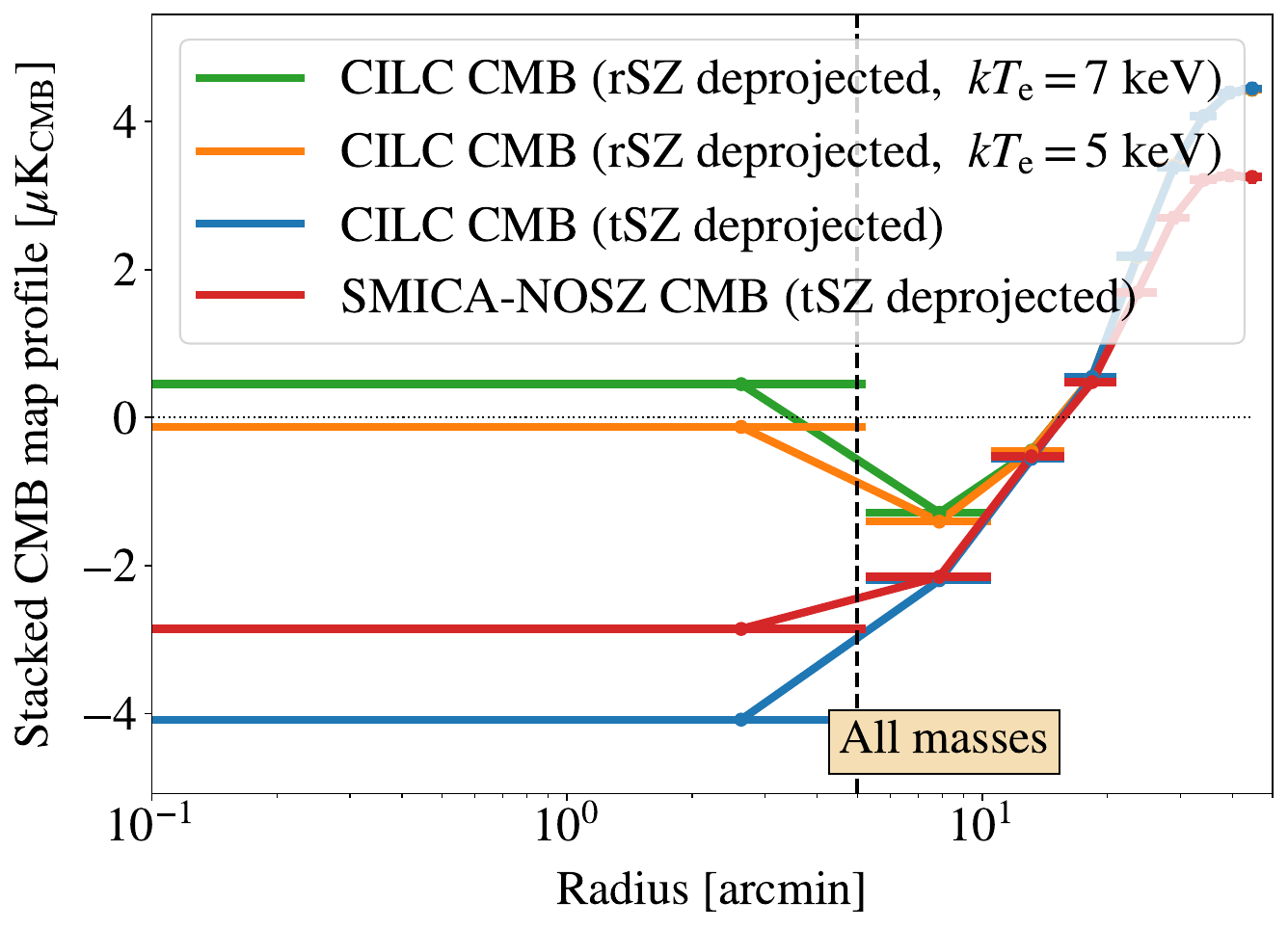}
      \includegraphics[width=0.85\columnwidth]{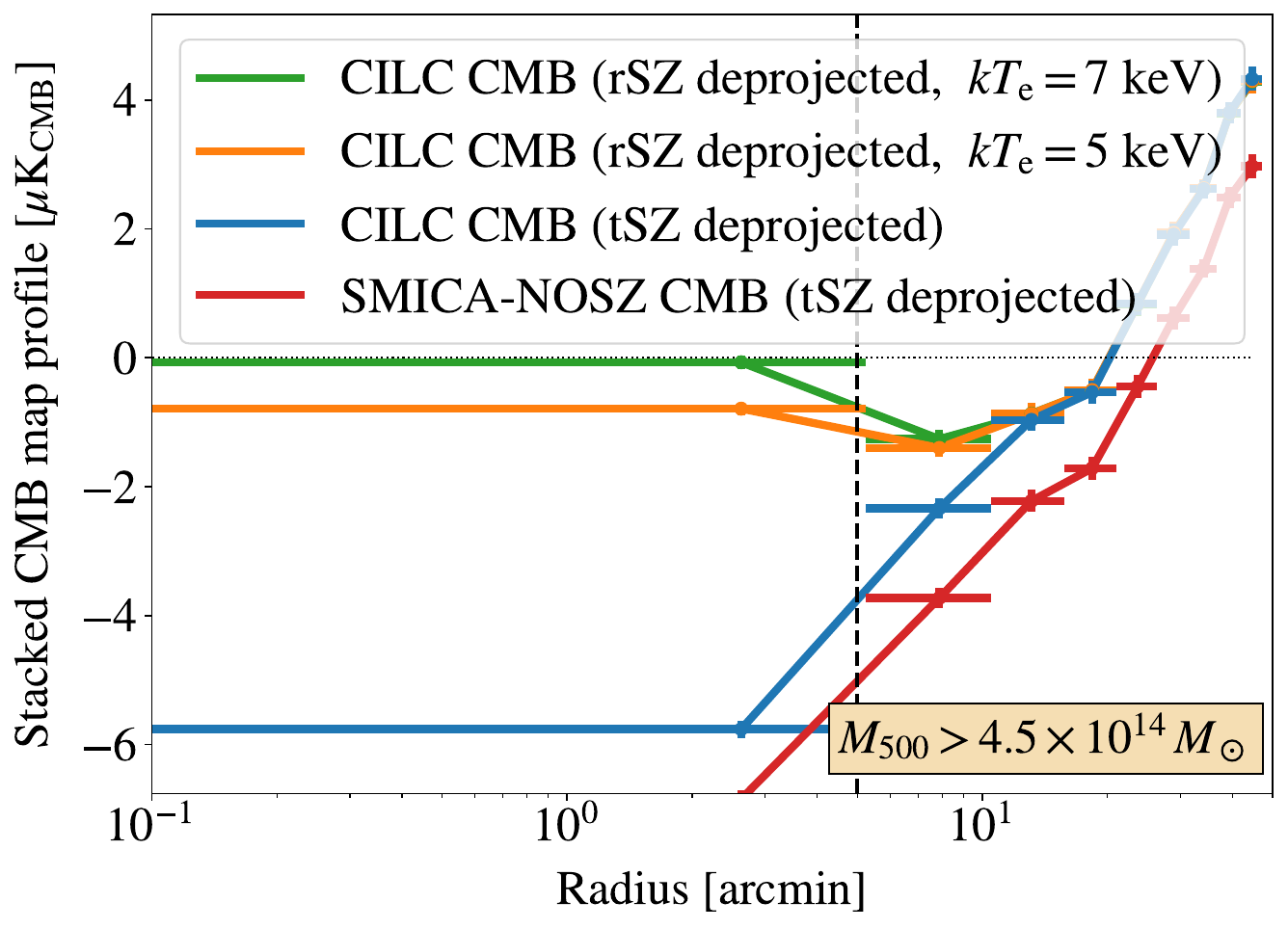}
       \caption{Profiles of \planck\ CMB maps stacked on \planck\ PSZ1 MMF1 clusters, either all of them (\emph{top}) or the most massive ones (\emph{bottom}), after tSZ deprojection with either SMICA (\emph{red}; cf. bottom panel in Figures.~\ref{fig:cmb} and \ref{fig:cmb_masscut}) or CILC (\emph{blue}; cf. middle panel in Figures.~\ref{fig:cmb} and \ref{fig:cmb_masscut}), and after rSZ deprojection assuming $T_{\rm e}=5$\,keV (\emph{orange}; cf. top panel in Figures.~\ref{fig:cmb} and \ref{fig:cmb_masscut}). Accounting for the relativistic SZ correction due to the average temperature $T_{\rm e} = 5$\,keV of the clusters results in the effective elimination of residual contamination from clusters in the CMB map within the $5'$ radius outlined by the vertical dashed line.}
    \label{fig:cmb_profile}
\end{figure}

 \begin{table}
    \centering
     \caption{Integrated flux in $\mu\text{K}_{\rm CMB}$ units of the stacked \planck\ CMB maps within the central circular aperture of radius $4.5'$ (outlined circle in Fig.~\ref{fig:cmb}), normalized by the aperture area. When stacking the \planck\ CMB maps on all 795 PSZ1 MMF1 clusters outside the Galactic mask, the minimum flux value is obtained by deprojecting the full rSZ emission at $T_{\rm e} = 5$\,keV.  When stacking the \planck\ CMB maps on  the 372 most massive clusters with $M_{500} > 4.5 \times 10^{14}\, M_\odot$ only, the minimum flux value is obtained by deprojecting the full rSZ emission at $T_{\rm e} = 7$\,keV.}
     \label{tab:cmb_flux}
     \begin{tabular}{lcc}
        \hline
                  Stacked CMB maps     & Flux & Flux \\
                       &  (all clusters) &   (massive clusters) \\
                   \hline
                   NILC & $-23.33 \pm 0.20$ & $-27.84 \pm 0.29$ \\
                    \hline
         SMICA-noSZ (tSZ-free) & $-2.87 \pm 0.12$ & $-7.22 \pm 0.23$  \\
         CILC (tSZ-free)  &$-4.23 \pm 0.13$ & $-6.19 \pm 0.23$\\
          \hline
         CILC (rSZ-free, $2$\,keV) &$-0.84 \pm 0.12$ & $-1.90 \pm 0.23$  \\
         CILC (rSZ-free, $3$\,keV) &$-0.51 \pm 0.12$ & $-1.49 \pm 0.23$  \\
         CILC (rSZ-free, $4$\,keV) &$-0.19 \pm 0.12$ & $-1.09 \pm 0.22$ \\
         CILC (rSZ-free, $5$\,keV) & $0.13 \pm 0.12$ &  $-0.69 \pm 0.22$  \\
         CILC (rSZ-free, $6$\,keV) &$0.45 \pm 0.12$ & $-0.29 \pm 0.22$  \\
         CILC (rSZ-free, $7$\,keV) &$0.76 \pm 0.12$ & $0.11 \pm 0.22$  \\
         CILC (rSZ-free, $8$\,keV) &$1.08 \pm 0.12$ & $0.50 \pm 0.22$  \\
         CILC (rSZ-free, $9$\,keV) &$1.39 \pm 0.12$ & $1.28 \pm 0.22$  \\
         CILC (rSZ-free, $10$\,keV) &$1.70 \pm 0.12$ &$9.74 \pm 0.22$  \\
        \hline
     \end{tabular}
    \end{table}

\begin{figure}
    \centering
    \includegraphics[width=0.8\columnwidth]{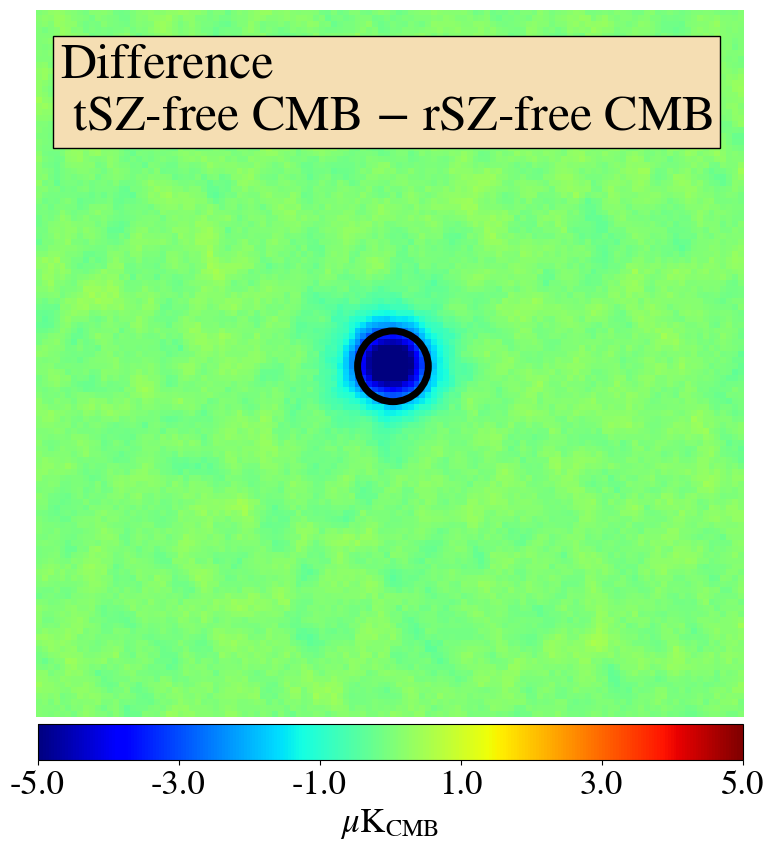}\\
     \includegraphics[width=0.8\columnwidth]{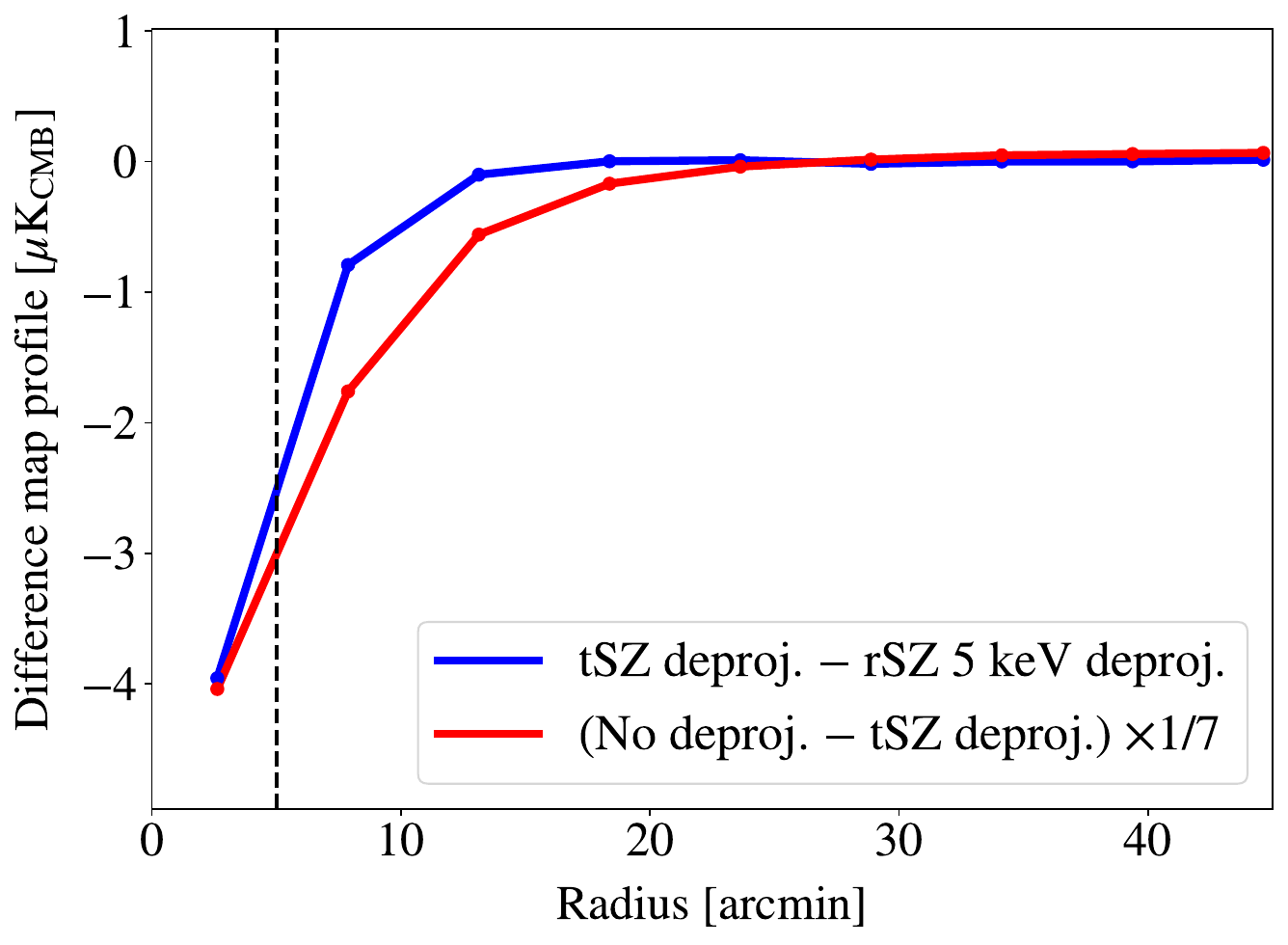}
    \caption{\emph{Top}: Difference between the stacked tSZ-free  \planck\ CILC CMB map (cf. middle panel in Figure~\ref{fig:cmb}) and the stacked rSZ-free  \planck\ CILC CMB map for $T_{\rm e}=5$\,keV (cf. top panel in Figure~\ref{fig:cmb}). The dimensions of the patch are $1.5^\circ\times 1.5^\circ$. \emph{Bottom}: Profile of the difference map shown on top (blue line).  This difference reveals the residual SZ contamination in tSZ-free \planck\ CMB maps when the relativistic SZ correction, stemming from the average temperature $T_{\rm e}=5$\,keV of the galaxy clusters, is overlooked. For comparison, the profile of the difference between the stacked tSZ-contaminated \planck\ NILC CMB map and the stacked tSZ-free \planck\ CILC CMB map \citep[see figure~2 in][]{Chen2018} is overlaid (red line), rescaled by a factor of $1/7$.}
    \label{fig:diff}
\end{figure}

Figure~\ref{fig:cmb_profile} presents the profiles of the stacked \planck\ CMB maps shown in Figs.~\ref{fig:cmb} and \ref{fig:cmb_masscut}, computed by integrating the signal within consecutive rings of $5'$ width. Accounting for the relativistic SZ correction with a temperature of $T_{\rm e} = 5$\,keV (orange line in top panel) is required to fully deproject the tSZ effect from the CMB map, eliminating the residual cluster contamination that persists in the non-relativistic tSZ-free CMB maps (blue and red lines). When stacking the CMB maps on the most massive \planck\ clusters, the most effective deprojection of rSZ contamination occurs when an average temperature of $T_{\rm e} = 7$\,keV is assumed (green line in bottom panel).

For completeness, Fig.~\ref{fig:diff} shows the difference between the two \planck\ CILC CMB maps (tSZ-free and rSZ free) from the middle and top panels of Fig.~\ref{fig:cmb}, along with the associated profile (blue line). This comparison provides further insight into the extent of residual SZ contamination that remains when the relativistic SZ temperature of $T_{\rm e} \simeq 5$\,keV for the \textit{Planck} galaxy cluster sample is omitted. This can also be compared to the extent of residual SZ contamination when none of the thermal SZ is deprojected, as shown by the red profile, which represents the difference between the tSZ-contaminated \planck\ NILC CMB map and the tSZ-free \planck\ CILC CMB map \citep[figure~2 from][]{Chen2018}. The residual tSZ contamination profile (red line) is roughly seven times more intense than the residual rSZ contamination profile (blue line), which exhibits a steeper radial profile attributed to the modulation of the $y$-profile by the relativistic electron temperature profile.

\subsection{Estimation from semi-analytical scaling relation}
\label{subsec:scaling_results} 

As an additional validation, we estimated the electron gas temperature of the \planck\ clusters based on the integrated Compton parameter values, $Y_{500}$, from the PSZ1 catalogue by applying the semi-analytical $Y$-$T$ scaling relation from \cite{Lee2022}:
\begin{equation}
\label{eq:scaling}
T = E(z)^{2/5} A(z) \left ( \frac{ Y }{ 10^{-5} } \right)^{B(z)}\, \text{keV}\,,
\end{equation}
where $E(z)=\sqrt{\Omega_m\left(1+z\right)^3+\Omega_k\left(1+z\right)^2+\Omega_\Lambda}$ is the ratio of the Hubble parameter at redshift $z$ to its present value $H_0$, with $\Omega_m$, $\Omega_k$, and $\Omega_\Lambda$ representing the total matter density, spatial curvature density, and dark energy density, respectively, as determined by the \Planck\ 2018 $\Lambda\text{CDM}$ best-fit model \citep{planck2016-l01}. The scaling parameters $A(z)$ and $B(z)$ are taken from the table~8 of \cite{Lee2022}. Since these parameters vary only slightly with redshift, they were linearly interpolated for the different redshifts in the cluster catalogue based on the representative values provided at five specific redshifts in \cite{Lee2022}. The scaling relation in Eq.~\eqref{eq:scaling} is particularly useful in the low signal-to-noise regime of \planck, where the relativistic SZ effect cannot be measured on individual clusters.

\begin{figure}
    \centering
    \includegraphics[width=0.8\columnwidth]{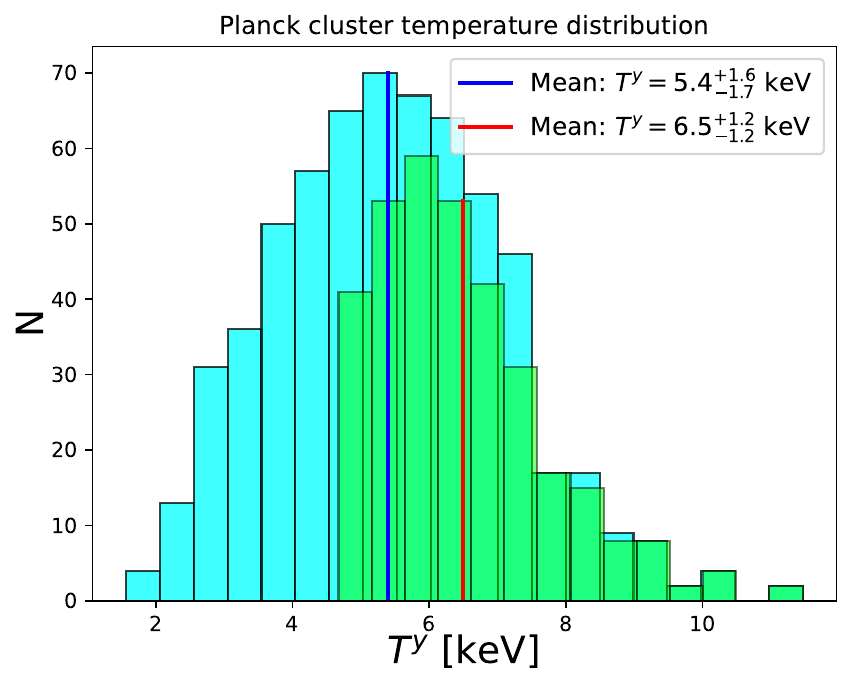}~
        \caption{Temperature distributions for the full \planck\ PSZ1 MMF1 cluster sample (cyan) and the subsample of the most massive clusters (green), derived from the semi-analytical scaling relation in Eq.~\eqref{eq:scaling}.}
    \label{fig:temphist}
\end{figure}

In light of the findings by \cite{Remazeilles2019} indicating about $8\%$ negative bias in the \planck\ Compton $y$-parameter (and hence the integrated flux $Y$) arising from the disregard of the relativistic SZ correction at $T_{\rm e} \simeq 5$\,keV in \planck\ SZ analyses, we  rescaled the measured $Y_{500}$ values from the \planck\ cluster catalogue by a factor of $1.08$ in the scaling relation of Eq.~\eqref{eq:scaling} before deriving the temperature. The resulting temperature distribution for the \planck\ clusters is plotted in Fig.~\ref{fig:temphist}, from which we derive an average temperature of $T^y \simeq 5.4^{+1.6}_{-1.7}$\,keV (68\% C.L.) for the full cluster sample. This result is in agreement with independent estimates from \cite{Zubeldia2024} and \cite{Erler2018}. For the subsample of the most massive clusters, with masses $M_{500} > 4.5\times 10^{14}M_\odot$, we obtain an average temperature of $T^y \simeq 6.5^{+1.2}_{-1.2}$\,keV (68\% C.L.) using the same scaling relation. These semi-analytical estimates are broadly consistent with the results from our map-based analysis presented above.

\section{Conclusions}
\label{sec:conclusions} 

In this work, we performed a non-parametric, map-based component separation analysis of the \planck\ data to map the relativistic SZ effect and estimate the average temperature of \planck\ galaxy clusters. This represents the first application of the method originally developed in \cite{Remazeilles2020} to real observational data.

We successfully generated the first rSZ first-order moment maps from \planck\ data. These maps, representing the electron-temperature-modulated Compton parameter field, $y(T_{\rm e}-\bar{T}_{\rm e})$, provide a novel approach to estimating the average temperature of the \planck\ clusters based solely on the rSZ effect. By mapping the $y(T_{\rm e}-\bar{T}_{\rm e})$ field for various pivot temperatures, $\bar{T}_{\rm e}$, and stacking the resulting maps on the positions of galaxy clusters, we inferred the average cluster temperature to be around $T_{\rm e} \simeq 5$\,keV, as indicated by the vanishing signal for $\bar{T}_{\rm e} = 5$\,keV. Furthermore, taking the ratio of the rSZ first-moment flux, $y(T_{\rm e}-\bar{T}_{\rm e})$, to the zeroth-moment flux, $y$, at $\bar{T}_{\rm e}=5$\,keV allowed us to estimate an average temperature of $T_{\rm e} = 4.9 \pm 2.6$\,keV for \planck\ clusters. By focusing on a subsample of the most massive clusters, we derived a higher average temperature of $T_{\rm e} = 7.3 \pm 2.1$\,keV using the same method.

In addition, we revised the tSZ deprojection process in \planck\ CMB maps, which had previously assumed a non-relativistic tSZ effect, by incorporating the average electron temperature into the tSZ SED model used for deprojection. This enabled us to produce rSZ-deprojected \planck\ CMB maps for different electron temperatures. By stacking on galaxy clusters, we demonstrated that residual signals from galaxy clusters in earlier tSZ-deprojected \planck\ CMB maps could be eliminated in the rSZ-deprojected maps, provided the average cluster temperature used for deprojection was $T_{\rm e} = 5$\,keV, corroborating our findings from the rSZ first-order moment maps.

We expect that our results can be further refined with higher angular resolution and sensitivity, for example by integrating data from both \planck\ and the Atacama Cosmology Telescope \citep[ACT --][]{ACT2020} into the map-based component separation pipeline. However, it is important to note that residual cosmic infrared background (CIB) contamination could affect the present analysis, as stacking on clusters also involves stacking on galaxies within those clusters that contribute to the CIB. This may introduce a slight bias in the inferred cluster temperature, which may be non-negligible with increased sensitivity and resolution. The consistency of our results with those from analytic estimates in Sect.~\ref{subsec:scaling_results} suggest that this effect may not be as important (see Appendix~\ref{sec:app2} for further discussion). In \cite{Zubeldia2024}, it was also argued that the CIB contamination in SZ cluster detection may not be as dramatic; however, this will have to be studied more carefully.

Upcoming and proposed CMB experiments, such as LiteBIRD \citep{LB2022,LB_SZ2024}, the Simons Observatory \citep{SO2019}, CMB-Stage 4 \citep{CMBS4-2019}, PICO \citep{PICO2019}, or a combination of these, are expected to provide the necessary sensitivity and angular resolution to conduct a similar analysis without relying on stacking \citep[see][]{Remazeilles2020}. This would allow for the measurement of electron temperatures for individual clusters and potentially establish the rSZ temperature field, $T_{\rm e}$, as a novel cosmological probe, complementing the Compton $y$-parameter field.

\section*{Acknowledgements}
The authors thank Will Coulton for fruitful discussion and comments.
We also thank the anonymous referee for their valuable suggestions.
M.R. acknowledges the support of the Spanish Ministry of Science and Innovation through projects~PID2022-139223OB-C21 and PID2022-140670NA-I00, funded by the Spanish MCIN/AEI/10.13039/501100011033/FEDER, UE.
J.C. was supported by the ERC Consolidator Grant \emph{CMBSPEC} (No. 725456) and the Royal Society as a Royal Society University Research Fellow at the University of Manchester, UK (No. URF/R/191023). 
The presented results are based on observations obtained with \planck,\footnote{http://www.esa.int/Planck} an ESA science mission with instruments and contributions directly funded by ESA Member States, NASA, and Canada. Some of the results in this paper have been derived using the \texttt{healpy} \citep{Zonca2019} and \textit{HEALPix} \citep{gorski2005} packages.

\section*{Data Availability}

The \planck\ rSZ-free CMB maps, the \planck\ rSZ $y$-maps, and the \planck\ rSZ first-order moment maps generated in this article for various pivot temperatures will be made available upon reasonable request to the authors.
The \planck\ PR2 data, the \planck\ SMICA-noSZ CMB map, and the \planck\ GAL80 Galactic mask can all be accessed via the Planck Legacy Archive.


\bibliographystyle{mnras}
\bibliography{main.bib,Planck_bib}



\appendix

\section{High-pass filtering and random stacking of relativistic SZ moment maps}\label{sec:app1}

As shown in the upper panel of Fig.~\ref{fig:moment_profile}, the stacked cluster signal from the $y(T_{\rm e}-\bar{T}_{\rm e})$ maps sits atop a large-scale negative residual that extends well beyond the cluster size when stacking all \planck\ clusters. However, no such residual is observed when stacking only the most massive clusters, as seen in the lower panel of Fig.~\ref{fig:moment_profile}. This suggests that the residual is uncorrelated with clusters and likely originates from residual Galactic contamination in the $y(T_{\rm e}-\bar{T}_{\rm e})$ map, which accumulates more significantly when stacking all galaxy clusters rather than a smaller subset.

\begin{figure}
    \centering
    \includegraphics[width=\columnwidth]{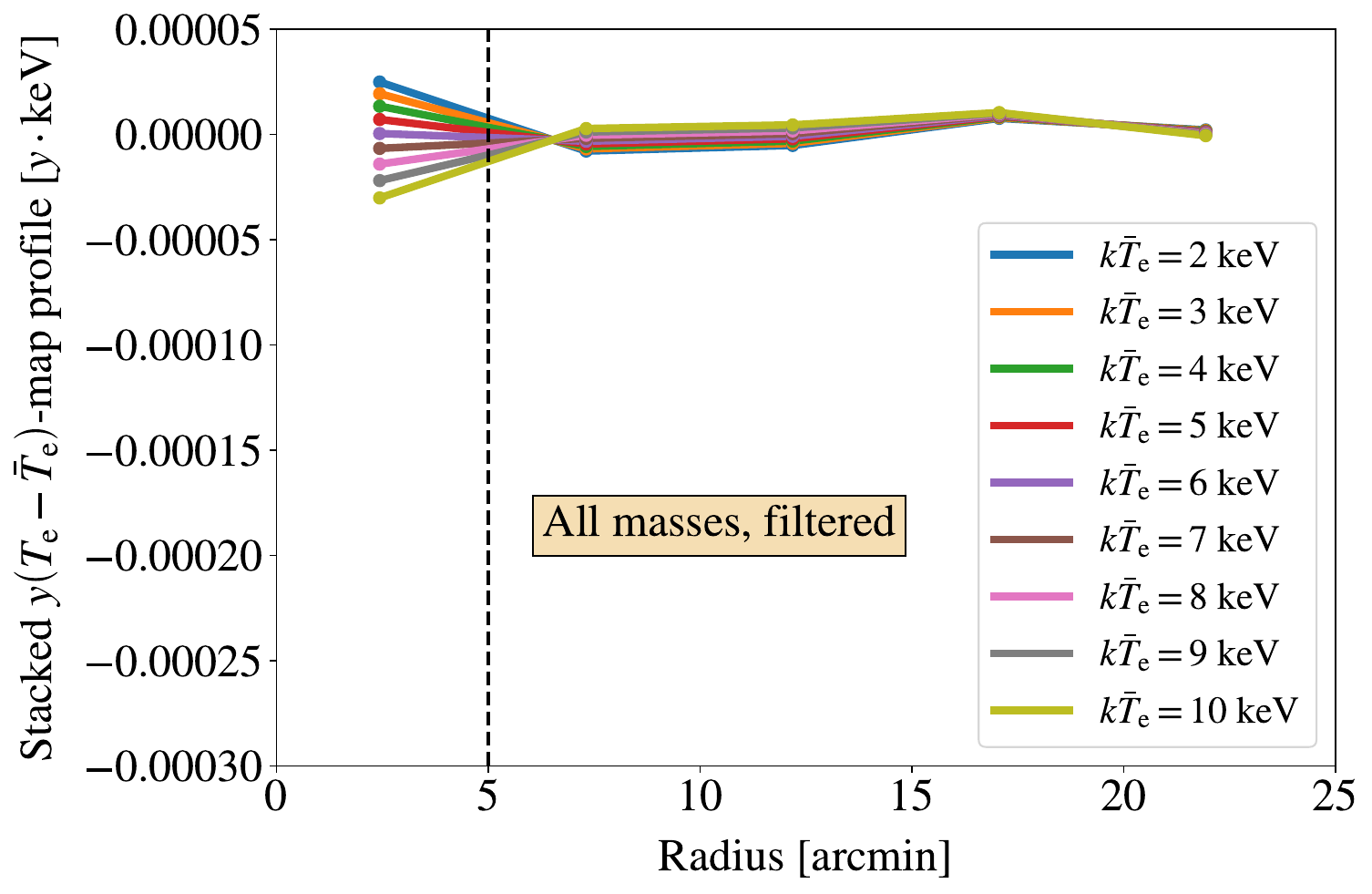}\\
    \includegraphics[width=\columnwidth]{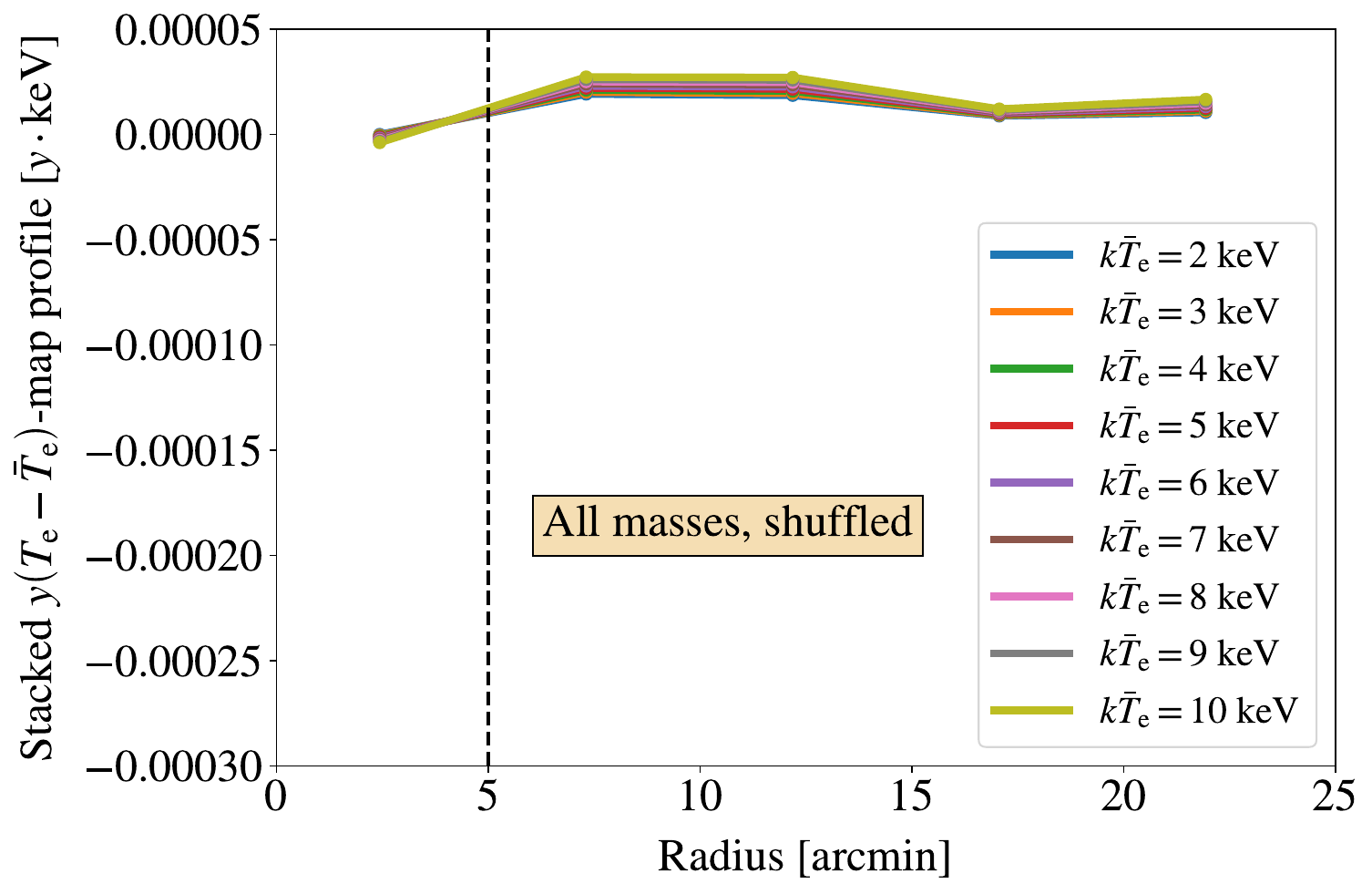}
    \caption{\emph{Top}: Same as the upper panel of Fig.~\ref{fig:moment_profile}, but after applying high-pass filtering to the $y(T_{\rm e}-\bar{T}_{\rm e})$ maps. \emph{Bottom}:  Same as the upper panel of Fig.~\ref{fig:moment_profile}, but with the $y(T_{\rm e}-\bar{T}_{\rm e})$ maps stacked at random positions by reshuffling the Galactic longitudes and latitudes of the cluster catalogue.}
    \label{fig:hp_moment_profile}
\end{figure}

To remove this large-scale residual, we can apply a high-pass filter in harmonic space, given by
\begin{align}
\label{eq:highpass}
f(\ell) = \left[ 1 + e^{ - c_1 \left( {\ell}/{\ell_c} - c_2 \right) } \right]^{-1}\,,
\end{align}
where $\ell_c = 850$, $c_1=40$, and $c_2=1$. This filtering effectively eliminates the large-scale residual for the full cluster sample stacking, as demonstrated in Fig.~\ref{fig:hp_moment_profile} (upper panel). Importantly, the characteristic slope variation of the profile with $\bar{T}_{\rm e}$ remains unchanged after high-pass filtering, confirming that the residual was independent of the cluster rSZ signal. In contrast, when stacking the $y(T_{\rm e}-\bar{T}_{\rm e})$ maps at random positions for various $\bar{T}_{\rm e}$ values, as shown in the lower panel of Fig.~\ref{fig:hp_moment_profile}, the characteristic slope variation associated with the cluster rSZ signal disappears.

\section{CIB deprojection in stacked CMB maps}\label{sec:app2}

 Deprojecting the relativistic SZ effect using the appropriate average electron temperature effectively removes the residual cluster signal from the stacked \planck\ CMB map (top panels of Figs.~\ref{fig:cmb} and~\ref{fig:cmb_masscut}), in contrast to the tSZ-free CMB map where a negative cluster residual remains after stacking due to the deprojection of only the non-relativistic SZ effect (middle panels of Figs.~\ref{fig:cmb} and~\ref{fig:cmb_masscut}). This contrast is further emphasized in Fig.~\ref{fig:diff}, which shows the difference between the tSZ-free and rSZ-free CMB maps.
 
Since stacking on clusters also involves stacking on the galaxies within them, the CIB could act as a contaminant to the rSZ effect. However, its variance is minimized through CILC, since its spectral signature is distinct from that of the CMB and rSZ. To strengthen our argument that the persisting negative cluster residual in the stacked tSZ-free CMB map (Fig.~\ref{fig:diff}) originates primarily from the relativistic SZ correction rather than the CIB contamination, we generate tSZ- and CIB-free CMB maps by deprojecting the non-relativistic tSZ effect along with the zeroth- and first-order moments of the CIB, instead of the rSZ effect.
 
According to \cite{planck2013-pip56}, dusty star-forming galaxies contributing to the CIB from redshift $\bar{z}$ emit as modified blackbodies with redshifted temperature:
\begin{align}
\label{eq:mbb}
I_\nu(\bar{z},\bar{\beta},\bar{T}) = \left[ \nu\left(1+\bar{z}\right)\right]^{\,\bar{\beta}}B_{\nu(1+\bar{z})}\left( \bar{T}=T_0\left(1+\bar{z}\right)^\alpha\right)\,,
\end{align} 
where $\bar{\beta} \simeq 1.75$, $T_0 \simeq 24.4$\,K, $\alpha \simeq 0.36$, and $B_\nu(T)$ is Planck's blackbody function.  Since we are interested in CIB emission from galaxies within \planck\ clusters, we adopt a fiducial redshift of ${\bar{z}=0.2}$, which represents the mean redshift of the \planck\ cluster sample.
As the observed CIB emission at each frequency receives contributions from multiple modified blackbodies along and across the lines of sight, its overall spectral signature can deviate from a modified blackbody. These deviations can be modelled using higher-order moments of the spectral parameters as \citep{chluba2017}:
\begin{align}
\label{eq:cib}
I^{\rm CIB}_\nu(\beta,T) &= I_\nu(\bar{z},\bar{\beta},\bar{T}) + \left(\beta - \bar{\beta}\right)\frac{\partial I_\nu}{\partial\bar{\beta}}(\bar{z},\bar{\beta},\bar{T}) \cr
& +  \left(T - \bar{T}\right)\frac{\partial I_\nu}{\partial\bar{T}}(\bar{z},\bar{\beta},\bar{T}) + \mathcal{O}(\beta^2,T^2)\,.
\end{align}

Over the \planck\ frequency range, the CIB spectrum is roughly consistent with a power law and is primarily dependent on the spectral index. Therefore, we focus on deprojecting the moments of the CIB spectral index, although tests confirmed that deprojecting CIB temperature moments does not affect our conclusions. In addition to deprojecting the tSZ effect with CILC, we thus deproject the zeroth- and first-order moments of the CIB, whose respective SEDs are given by $I_\nu(\bar{z},\bar{\beta},\bar{T})$ (Eq.~\ref{eq:mbb}) and $\frac{\partial I_\nu}{\partial\bar{\beta}}(\bar{z},\bar{\beta},\bar{T}) = \log(\nu)I_\nu(\bar{z},\bar{\beta},\bar{T})$.

The results are summarized in Table~\ref{tab:cib_deproj} for the subsample of massive \planck\ clusters. When deprojecting CIB instead of rSZ, the negative residual flux persists at cluster positions in the CMB map, 
even when varying the CIB spectral index pivot value over a broad range from $\bar{\beta} = 1$ to $2.5$. This provides further evidence that the signal observed in Fig.~\ref{fig:diff} is attributable to the rSZ effect rather than the CIB. Moreover, as shown in Table~\ref{tab:cib_deproj}, the residual flux in the tSZ- and CIB-free CMB maps remains consistently negative and stable across different CIB spectral index pivots. This behaviour contrasts sharply with the characteristic signature of the rSZ signal, as seen in Table~\ref{tab:cmb_flux}, where the residual flux in the CMB map transitions from negative to positive values depending on the pivot cluster temperature $\bar{T}_{\rm e}$, with a null value at the actual average cluster temperature.

 \begin{table}
    \centering
     \caption{Integrated flux (in $\mu\text{K}_{\rm CMB}$ units) of the stacked \planck\ tSZ- and CIB-free CMB maps within a central circular aperture of radius $4.5'$ (outlined in Fig.~\ref{fig:cmb}), normalized by the aperture area. Unlike the deprojection of the rSZ effect (Table~\ref{tab:cmb_flux}), deprojecting the CIB leaves a persistent negative residual flux at cluster positions in the CMB map, regardless of the chosen pivot value for the CIB spectral index.}   
     \label{tab:cib_deproj}
     \begin{tabular}{lc}
        \hline
                Stacked CMB maps  & Flux \\
                       &   (massive clusters) \\
          \hline
         CILC (tSZ-free, CIB-free, $\bar{\beta} = 1$) & $-3.84 \pm 0.22$  \\
         CILC (tSZ-free, CIB-free, $\bar{\beta} = 1.75$) & $-3.83 \pm 0.22$  \\
         CILC (tSZ-free, CIB-free, $\bar{\beta} = 2.5$) & $-3.83 \pm 0.22$  \\         
         \hline
     \end{tabular}
    \end{table}

\bsp	
\label{lastpage}
\end{document}